\documentclass[aps,10pt,prd,nofootinbib,superscriptaddress,onecolumn]{revtex4-2}
\pdfoutput=1

\usepackage[utf8]{inputenc}
\usepackage{geometry}
\usepackage{mathtools}
\usepackage{amsfonts}
\usepackage{dsfont}
\usepackage{mathrsfs}
\usepackage{bbm}
\usepackage[normalem]{ulem}
\usepackage{slashed}
\usepackage{tensor}

\usepackage{pifont}

\usepackage{graphicx}
\usepackage{array}

\usepackage{placeins}
\usepackage{makecell}
\usepackage{float}

\usepackage{xspace}
\usepackage{xfrac}
\usepackage{hyperref}
\usepackage[nameinlink]{cleveref}
\usepackage{appendix}
\usepackage{units}

\usepackage{xifthen}
\usepackage{booktabs}
\usepackage{multirow}
\usepackage[dvipsnames]{xcolor}
\hypersetup{
	colorlinks,
	linkcolor={red!75!black},
	citecolor={blue!75!black},
	urlcolor={blue!75!black},
	pdftitle={Juggling with Tensor Bases in Functional Approaches},
	pdfauthor={Braun, Geissel, Pawlowski, Sattler, Wink},
}
\usepackage{enumerate}

\usepackage{mmacells}

\mmaDefineMathReplacement[≤]{<=}{\leq}
\mmaDefineMathReplacement[≥]{>=}{\geq}
\mmaDefineMathReplacement[≠]{!=}{\neq}
\mmaDefineMathReplacement[→]{->}{\to}[2]
\mmaDefineMathReplacement[⧴]{:>}{:\hspace{-.2em}\to}[2]
\mmaDefineMathReplacement{∉}{\notin}
\mmaDefineMathReplacement{∞}{\infty}

\mmaDefineMathReplacement{𝕕}{\mathbbm{d}}
\mmaSet{
	leftmargin=3.5em,
	morefv={gobble=0},
	linklocaluri=mma/symbol/definition:#1,
	morecellgraphics={yoffset=1.9ex},
	labelsep=0.5em
}

\newcommand{\innerproduct}[2]{\left\langle #1, #2 \right\rangle}
\newcommand{\imag}{\text{i}}
\newcommand{\norm}[1]{\left\lVert#1\right\rVert}

\newcommand{\tinytext}[1]{\text{\tiny{#1}}}
\newcommand{\TensorBases}{\texttt{TensorBases}\xspace}

\newcommand{\LEGO}{LEGO\textsuperscript{\textregistered}}

\newcommand{\orcid}[1]{\href{https://orcid.org/#1}{\includegraphics[height=1.9ex,width=1.9ex]{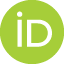}}}

\newcommand{\gettitle}{Juggling with Tensor Bases in Functional Approaches}

\newcommand{\getHeidelbergAffiliation}{\affiliation{Institut f{\"u}r Theoretische Physik, Universit{\"a}t Heidelberg, Philosophenweg 16, 69120 Heidelberg, Germany}}
\newcommand{\getDarmstadtAffiliation}{\affiliation{Institut f\"ur Kernphysik (Theoriezentrum), Technische Universit\"at Darmstadt, 64289 Darmstadt, Germany}}
\newcommand{\getEMMIAffiliation}{\affiliation{ExtreMe Matter Institute EMMI, GSI, Planckstr. 1, 64291 Darmstadt, Germany}}

\begin{document}

\title{\gettitle}
\author{Jens Braun \orcid{0000-0003-4655-9072}\,}
\getDarmstadtAffiliation
\getEMMIAffiliation
\author{Andreas Geißel \orcid{0009-0007-9283-4211}\,}
\getDarmstadtAffiliation
\author{Jan M. Pawlowski \orcid{0000-0003-0003-7180}\,}
\getHeidelbergAffiliation
\getEMMIAffiliation
\author{Franz R. Sattler \orcid{0000-0003-1744-9456}\,}
\getHeidelbergAffiliation
\author{Nicolas Wink \orcid{0000-0002-2811-454X}
}

\begin{abstract}

Systematic expansion schemes in functional approaches require the inclusion of higher order vertices. These vertices are expanded in independent tensor bases with a rapidly increasing number of basis elements. 
Amongst the related tasks are the construction of bases and projection operators, the importance ordering of their elements, and the optimisation of such tensor bases, as well as an analysis of their regularity in momentum space.
We present progress in all these directions and introduce the Mathematica package \href{https://github.com/satfra/TensorBases.git}{\TensorBases} designed for the aforementioned tasks.

\end{abstract}

\maketitle

\section{Introduction}
\label{sec:introduction}

Functional approaches are one of the few truly non-perturbative approaches that allow for a comprehensive access to the physics of complex strongly correlated systems. A specifically appealing property is the incredibly simple access to even semi-quantitative results. While this can be called a universal feature, many phenomena require a fully quantitative access for even capturing the qualitative properties of the system at hand. Specific examples are regimes in the phase structure of a given system with competing orders. This includes systems ranging from condensed matter physics to high energy physics. A less intricate but related phenomenon is the change of ordering phenomena and the respective dynamics. All these transitions are difficult to access as they typically involve a fully intertwined dynamics of both regimes. For reviews on functional approaches related to the present work on tensor bases see, e.g.~\cite{Fu:2022gou, Pawlowski:2020qer, Dupuis:2020fhh, Platt_2013, Metzner:2011cw, Braun:2011pp, Gies:2006wv} for applications of the functional renormalisation group (fRG) to a broad range of systems, ranging from condensed matter systems over QCD to quantum gravity, \cite{Fischer:2018sdj, Eichmann:2016yit, Bashir:2012fs, Fischer:2006ub, Maris:2003vk, Alkofer:2000wg, Roberts:1994dr} for QCD-applications of Dyson-Schwinger equations (DSEs), and \cite{Klevansky:1992qe,Buballa:2003qv, Buballa:2014tba,Andersen:2014xxa} for low energy effective theories of QCD with Nambu-Jona-Lasinio (NJL)-type interactions. 

We shall use Quantum Chromodynamics (QCD) at finite density as an illustrative, and physically challenging and exciting, but not comprehensive example: at low densities, the low-energy regime of QCD is governed by the pion dynamics, and the respective quark--anti-quark scattering vertex is completely dominated by the resonant pseudo-scalar channel. In turn, at large densities, the diquark channel takes over the role of the dominant channel, see~\cite{Braun:2019aow} for an evaluation of this transition for two-flavour QCD. While this two-flavour case already appears involved, it still comes with significantly fewer competing channels than the phenomenologically more relevant case with $2+1$ flavours. 
The region between these two asymptotic regimes with fully coupled dynamics is indeed difficult to resolve, but it is particularly relevant, as it will be covered by upcoming heavy ion experiments. 

The resolution of this intricate dynamics calls for systematic expansion schemes in functional approaches, that combine both a well-developed theory for systematic error estimates on the conceptual level, as well as the computational capacity to achieve the numerical accuracy required for small systematic errors. 
In the past decade, much progress has been made in both directions and we refer to \cite{Ihssen:2024miv, Fu:2025hcm, Huber:2025kwy} for recent developments in functional approaches. The underlying expansion schemes build on the simple one- or two-loop diagrammatic structure of the functional relations for correlation functions of the theory at hand.
This leads to combinations of a vertex expansion, the accommodation of resonant degrees of freedom with emergent composites, as well as full scattering potentials of dynamical low-energy degrees of freedom.
The latter interactions are only important for low momenta and can be expanded in terms of momenta, measured in the low energy scales of the theory at hand. This is called the derivative expansion.
This leads to a combination of different sectors of the theory, that are only connected dynamically by a few diagrammatic ``anchor bolts". This modular structure facilitates the systematic error analysis, and the underlying principle has been baptised the \LEGO-principle~\cite{Ihssen:2024miv}. 

A major building block  of such an approach is the complete and optimised control of higher order scattering vertices. 
To begin with, this asks for the construction of complete tensor bases for a given vertex, but it does not stop there. 
Such bases are not unique, the Fierz ambiguity in the four-Fermi vertex being a prime example. Furthermore, in two-flavour QCD, a Fierz-complete basis has approximately 250 elements in the vacuum \cite{Eichmann:2015cra}, and this increases dramatically if, for example, switching on finite temperature or chemical potential, or going to 2+1 flavours. 
A specific task related to momentum-dependent tensors, is the choice of a regular basis. Evidently, we can shift momentum dependences from the dressings to the tensors and vice versa. However, this may cause irregularities or remove them. Moreover, we may have to face momentum-dependent singularities in the completeness relations of the basis. 
Leaving these intricacies aside, we may reduce the basis in the spirit of the derivative expansion mentioned above to the momentum-independent ones. In two-flavour QCD this leads us to a basis with 10 elements. As explained before, for small or vanishing densities, the pseudoscalar pion channel dominates the dynamics and we may restrict ourselves to only one or a few of the ten tensors. Then, the basis can be optimised such that the information in this specific incomplete set of tensors can be maximised.

The conceptual and computational intricacies briefly discussed above can be dealt with in a framework that allows to resolve the respective computational tasks. For this purpose, we have developed the Mathematica package \TensorBases which already comes with a set of different bases for the quark-gluon vertex, the four-quark vertex and several others. Moreover, the package is set up in such a way that it allows for a successive extension to further vertices with user-defined group structures.

The present work is organized as follows.
In \Cref{sec:general}, we give a short summary of the necessary linear algebra from a very general point of view and illustrate the main concept underlying our present work with a very simple example.
In \Cref{sec:inner_product}, we study the importance of a suitably chosen inner product that can be used for projection onto basis elements where fermions have to be treated with some additional care, as we shall show explicitly.
The choice of a particular basis and restrictions to relevant subspaces thereof are investigated in \Cref{sec:optimisation}, including a discussion of the optimization of tensor bases.
Eventually, the momentum structure of a tensor basis will be investigated in \Cref{sec:momenta}, with the quark-gluon vertex as a concrete example.
In \Cref{sec:fourFermiVertex}, we illustrate our approach using the four-quark vertex.
As the present work comes together with the Mathematica package \TensorBases, we present the installation steps as well as the basic usage of the package in \Cref{sec:package}.
Our conclusions can be found in \Cref{sec:conclusions}.

\section{Vector spaces, projectors and the metric}
\label{sec:general}

In this section, we discuss the necessity of a metric in the tensor space of a given vertex for the optimisation and expansion tasks in functional studies. To that end, we first consider the general case in \Cref{sec:Preliminaries} before we illustrate the use of the metric within a simple example in three-dimensional Euclidean space in \Cref{sec:simpleexpample}.

\subsection{Preliminaries}
\label{sec:Preliminaries} 

Let $V$ be a vector space and $\innerproduct{\cdot}{\cdot}$ an inner product on $V$, so that $(V,\,\innerproduct{\cdot}{\cdot})$ is an inner product space. 
The simplest example for such a space is  $\mathbb{R}^n$ with its canonical inner product, i.e. the standard scalar product. 
In the present work, we shall consider $V$ to be the space of all possible tensor structures of a specific multi-particle interaction that respect the symmetries of the theory under consideration.
For example, this may be the space of all possible four-fermion interactions in two-flavour QCD at finite chemical potential and temperature.

Let us take $n$ arbitrary vectors $\{e_i\}_{i=1\dotsc n}\subseteq V$. We can now check whether or not any of these vectors are linearly dependent and whether they form a basis.
If the set $\{e_i\}$ is complete, $\text{span}(\{e_i\}) = V$, and linearly independent, then the set $\{e_i\}$ is a basis.
Suppose that we have already checked the completeness, $\text{span}(\{e_i\}) = V$, but have not yet checked the linear independence. 
For this purpose, we define the metric
\begin{equation}\label{eq:ex1_metric}
	g_{ij} = \innerproduct{e_i}{e_j}
	\,.
\end{equation}
As indicated above, we can check the linear dependence of the basis at hand by calculating the determinant of~\labelcref{eq:ex1_metric}. 
If it is non-zero,
\begin{equation}\label{eq:ex1_det}
	\left|g \right|  \equiv \mathrm{det} g_{ij} \neq 0
	\,,
\end{equation}
the set $\{e_i\}$ is a basis of $V$. 
In case the determinant vanishes, the set of vectors is over-complete.
Consequently, we can always achieve a non-vanishing determinant of $g$ by simply eliminating vectors from $\{e_i\}$ while keeping the completeness relation $\text{span}(\{e_i\}) = V$. 
Eventually, due to \labelcref{eq:ex1_det}, $g_{ij}$ is invertible.
In this case, $g_ {ij}$ is the coordinate-representation of the metric tensor induced by the inner product.
We remark that the basis is called orthogonal if the metric $g_{ij}$ is diagonal and orthonormal if it is the unit matrix.

The decomposition of a vector $v \in V$ in terms of the basis vectors $\{e_i\}$ is given by
\begin{equation}\label{eq:ex1_decompose}
	v = v_i e_i
	\,,
\end{equation}
while the coefficients $v_i$ are given by
\begin{equation}\label{eq:ex1_decompose_expl}
	v_i = g^{-1}_{ij} \innerproduct{e_j}{v}
	\,.
\end{equation}
Hence, we conclude that \labelcref{eq:ex1_decompose_expl} is a straightforward prescription to perform projections onto basis elements of tensor bases that appear, e.g. in multi-particle scatterings in quantum field theory (QFT).
	
The procedure explained above is specifically simple if the vector space and its inner product are known and the entries of the metric are constant. However, for QFTs, it turns out that arranging for both properties is intricate and requires a careful discussion. 
Thus, we shall discuss the correct choice of an inner product in \Cref{sec:inner_product} and intricacies of momentum dependences in \Cref{sec:momenta}.
Nevertheless, it is useful to illustrate the above procedure with a very simple example where both caveats are absent.

\subsection{A simple example}
\label{sec:simpleexpample}

Let us begin with a simple example, demonstrating the setup given above in three-dimensional Euclidean space~$\mathbb{R}^3$.
Furthermore, let us choose the canonical basis and change it into some new basis.
This step is needed to illustrate the procedure from above.
The inner product in this case is simply given by the standard scalar product, i.e.
\begin{equation}\label{eq:ex1_inner_product_R3}
	\innerproduct{x}{y} = x_i y_i
	\,,
\end{equation}
where the summation over repeated indices is implied.
For the new rotated basis we use the basis vectors
\begin{equation}
	e_1 = \begin{pmatrix}
		-5\\ 3\\ 4
	\end{pmatrix}
	\quad
	e_2 = \begin{pmatrix}
		4\\ 2\\ 5
	\end{pmatrix}
	\quad
	e_3 = \begin{pmatrix}
		0\\ -2\\ -1
	\end{pmatrix}
	\,.
\end{equation}
Note that this basis is neither orthogonal nor normalised and thus we do not expect a diagonal metric.
Applying~\labelcref{eq:ex1_metric}, the metric can be evaluated to be
\begin{equation}
	g
	= \begin{pmatrix}
		50 & 6 & -10 \\
		6 & 45 & -9 \\
		-10 & -9 & 5
	\end{pmatrix}
	\,.
\end{equation}
Due to the symmetry of the inner product, the metric is symmetric as well.
The determinant turns out to be $|g| = 3600$. 
Finally, the inverse of the metric is given by
\begin{equation}
	g^{-1} = \begin{pmatrix}
		\frac{1}{25} & \frac{1}{60} & \frac{11}{100} \\[6pt]
		\frac{1}{60} & \frac{1}{24} & \frac{13}{120} \\[6pt]
		\frac{11}{100} & \frac{13}{120} & \frac{123}{200}
	\end{pmatrix}
	\,.
\end{equation}
To illustrate the projection prescription, let us consider the following vector
\begin{equation}
	d = \begin{pmatrix}
		1\\ 15\\ 1
	\end{pmatrix}
	\,.
\end{equation}
In an actual application, this vector may represent, for example, a diagram appearing on the right-hand side of a functional renormalisation group (fRG) equation or Dyson-Schwinger equation (DSE), which one would like to decompose into tensor basis components.

According to \labelcref{eq:ex1_decompose_expl}, the coefficients~$d_i^{e}$ are given by
\begin{equation}
	d_i^{e} = g^{-1}_{ij} \left\langle e_j, d \right\rangle
	\,,
\end{equation}
and the corresponding vector in the rotated basis yields
\begin{equation}
	d^{e} = \begin{pmatrix}
		-1\\ -1\\ -10
	\end{pmatrix}
	\,.
\end{equation}
The result can be verified by calculating $d_1^{e} \cdot e_1 + \ldots$ in canonical coordinates. Of course, this result could have been obtained in a much simpler manner by simply writing down the transition matrix between the two bases. 
We emphasise that the presented procedure only works if we know the inner product, without any reference to the original basis. 
We shall in the following consider this issue in the context of QFTs.

\section{Inner product, projection operators and basis construction}
\label{sec:inner_product}

The $n$-point vertices under consideration in functional approaches are typically the one-particle irreducible~(1PI) parts of the diagrammatic representation of correlation functions that constitute $n$-point scattering processes in a given QFT. 
They can be understood as expansion coefficients in a vertex expansion (in powers of the fields) of the respective 1PI effective action or related generating functionals. 
For our general discussion we introduce the functional~$F[\Phi]$, that depends on the superfield~$\Phi$. Typically, the superfield contains all fundamental fields but it also may contain (emergent) composite fields such as local fermionic bilinears or even products of fields at different space-time points. Moreover, it carries a dependence on the fundamental parameters of the underlying QFT, such as couplings and masses, as well as further scales, such as the infrared cutoff scale~$k$ in the fRG approach.
In the latter case, the master functional $F[\Phi]$ is the scale-dependent effective action, $F[\Phi] = \Gamma_k[\Phi]$. This functional is also the master object in related functional approaches such as the Dyson-Schwinger approach and $n$PI-approaches. In the latter case the superfield contains also the propagator (2PI) or higher-order vertices as separate fields. For a brief explanation of the notation used in the present work, especially the superfield formalism, we refer the reader to \Cref{app:notation}.

In a vertex expansion, $F[\Phi]$ is expanded in products of component fields $\Phi_i$ up to arbitrary order and the expansion coefficients are only constrained by the symmetries of the theory under consideration. Note that in the following we will only consider fields that become spatially constant on the quantum equations of motion. If one also takes into account spatially varying fields with finite expectation values, as necessary for the description of inhomogeneous phases, see Ref.~\cite{Buballa:2014tba} for a review, the associated tensor basis may be greatly enlarged. Thus, in the following we will only consider fields $\Phi$ that are  homogeneous or vanishing on the equation of motion.

As a specific example we consider a product of $n$ fields. The space of allowed vertices, related to the allowed field operators of order $n$, is typically large and is spanned by some basis.
The object we have to consider now is $F_\alpha\left[\Phi\right]$ with the multi-index $\alpha$ with $|\alpha| = n$. The functional  $F_\alpha\left[\Phi\right]$ is the $n$-th derivative of a given master functional $F[\Phi]$, see 	\labelcref{eq:Falpha}. With a given tensor basis $\mathcal{B} = \lbrace \mathcal{O}^{(i)} \rbrace$ for vertices of order $n$ in the fields, the parametrisation in terms of the basis elements $\mathcal{O}^{(i)} $ and associated dressings $\lambda_{i,\mathcal{B}}$ is provided in \labelcref{eq:general_decomposition}. Here,~$i$  denotes the index of the basis element, while $\mathcal{B}$ contains the information about the operator basis such as the type of $n$-field operator. Note that we have not displayed the momentum dependence of the dressing for the sake of readability. 

Finally, in functional approaches, we have to project such a general $n$-point functional onto the diagrammatic relations for the dressings $\lambda_{i,\mathcal{B}}$. Starting from $F_\alpha\left[\Phi\right]$, we aim to define a projection operator $\hat{P}_{\lambda_{i,\mathcal{B}}}^\alpha$ with the following property
\begin{equation}\label{eq:property_projection}
	\hat{P}_{\lambda_{i,\mathcal{B}}}^\alpha F_\alpha[\Phi] = \lambda_{i,\mathcal B}
	\,,
\end{equation}
where the summation over the multi-index $\alpha$ is tacitly assumed.
For the construction of the projection operator, we need to define a metric in operator space which naturally arises from an inner product. 
Let us consider the case where we operate on a space spanned by~$n$-fields~$\Phi^a$ that may be either complex- or Grassmann-valued.
For two elements $i,j$ of the set of tensors~$\mathcal{B} = \lbrace \mathcal{O}^{(i)}_\alpha \rbrace$, that span the specific space of $n$-field operators, e.g. four-fermion operators, we define the inner product via derivatives to be
\begin{equation}\label{eq:metric}
	g_{\mathcal B}^{ij} = \langle\mathcal{O}^{(i)},\mathcal{O}^{(j)}\rangle \equiv \mathcal{O}^{(i)}_{a_1,\dots,a_n}\frac{\delta}{\delta\Phi_{a_1}}\dots\frac{\delta}{\delta\Phi_{a_n}}
	\left[
	\mathcal{O}^{(j)}_{b_1,\dots,b_n}\Phi^{b_1}\dots\Phi^{b_n}\right]
	\,,
\end{equation}
where summation over repeated indices is tacitly assumed.
Note that the indices $i,j$ refer to the tensor elements of the operator space under consideration, while the index $\mathcal B$ indicates the field structure of the operator space. Then, the projection operator onto the dressing $\lambda_{i,\mathcal B}$ is given by
\begin{equation}
	\hat P_{\lambda_{i,\mathcal{B}}}^\alpha = \left(g^{-1}_{\mathcal{B}}\right)_{ij}\mathcal{O}^{(j),\alpha}
	\,.
\end{equation}
The required property \eqref{eq:property_projection} can be verified by inserting the definitions of the projection operator as well as of~$F_\alpha\left[\Phi\right]$.
Indeed, we find
\begin{equation}
	\langle \hat P_{\lambda_{i,\mathcal B}},F[\Phi] \rangle 
	= \hat P_{\lambda_{i,\mathcal B}}^\alpha,F_\alpha\left[\Phi\right]
	= \lambda_{i,\mathcal B}
	\,,
\end{equation}
where the inner product~$\langle \cdot , \cdot \rangle$ is that used for the definition of the metric in \labelcref{eq:metric}.

We would like to emphasise an important aspect regarding Grassmann-valued fields.
Crucially, the antisymmetric nature of Grassmann-valued fields is not contained in the tensors themselves, but is present in the inner product. The definition \labelcref{eq:scalarProduct} of the inner product respects this property. This concerns applications to four-fermion systems at all length scales, see, e.g. the reviews \cite{Pawlowski:2020qer, Dupuis:2020fhh} (gravity), 	\cite{Fu:2022gou, Pawlowski:2020qer, Dupuis:2020fhh, Braun:2011pp, Gies:2006wv, Eichmann:2016yit, Fischer:2018sdj, Ferreira:2023fva} (first-principles QCD), 
\cite{Klevansky:1992qe, Buballa:2003qv, Buballa:2014tba, Dupuis:2020fhh} (QCD low energy effective theories) and \cite{Dupuis:2020fhh, Platt_2013, Metzner:2011cw} (condensed matter ans statistical systems). For the latter systems we also refer to selected relevant works \cite{PhysRevB.79.085116, PhysRevB.108.235120}, where complete bases in condensed matter systems are discussed. 

As an explicit example we consider NJL-type low energy effective theories of QCD with four-fermion interactions (NJL-type theories), but the discussion below applies as well to all the examples mentioned above. The procedure described below has been previously introduced in similar form within the context of the fRG approach in \cite{Gehring:2015vja}. To begin with, we shall use that a  basis $\mathcal{B} = \lbrace \mathcal{O}_\alpha^{(i)} \rbrace$ for the four-fermion operators can be constructed from the tensor structures of fermionic currents $\mathcal{F}^{(i)}$. This will be discussed in detail in \Cref{sec:fourFermiVertex}. The respective basis elements $\mathcal{O}^{(i)}$ are given by 
\begin{equation}
	\mathcal{O}^{(i)}_\alpha  
	= \mathcal{O}^{(i)}_{a_1,a_2,a_3,a_4} 
	= \mathcal{F}^{(i)}_{a_1,a_2}\mathcal{F}^{(i)}_{a_3,a_4}\,,\qquad i =1,\dots,N_\textrm{4f}
	\,. 
	\label{eq:4FcurrentBasis}
\end{equation}
The full four-fermion operators are obtained by contracting the $\psi$-fields with the indices~$a_2$ and $a_4$, respectively, and the $\bar{\psi}$-fields with the indices $a_1$ and $a_3$, respectively,  to wit 
\begin{align} 
\left(\bar \psi^{a_1} 	\mathcal{F}^{(i)}_{a_1,a_2}\psi^{a_2}\right)\, \left( \bar \psi^{a_3} \mathcal{F}^{(i)}_{a_3,a_4}  \psi^{a_4} \right) \,.
\label{eq:CurrentSquared}
\end{align}
Note that the index $i$ in \labelcref{eq:4FcurrentBasis,eq:CurrentSquared} labels the tensor element of the four-fermion basis and hence is not summed over. This index specifies the tensor element of the basis of the four-fermion interactions. 
By inserting this definition into the definition of the metric \labelcref{eq:metric}, we find
\begin{align}\label{eq:scalarProduct}
	g_{\mathcal{B}}^{ij} = \langle\mathcal{O}^{(i)},\mathcal{O}^{(j)}\rangle 
	&=
	\mathcal{O}^{(i)}_{a_1,a_2,a_3,a_4}\
	\frac{\delta}{\delta\psi_{a_1}}\frac{\delta}{\delta\bar\psi_{a_2}} \frac{\delta}{\delta\psi_{a_3}}\frac{\delta}{\delta\bar\psi_{a_4}}
	\cdot
	\mathcal{O}^{(j)}_{b_1,b_2,b_3,b_4}\bar\psi^{b_1}\psi^{b_2}\bar\psi^{b_3}\psi^{b_4}
	\\[1ex]
	&= 2 \left(\mathcal{F}^{(i)}_{a_1,a_2}\mathcal{F}^{(j)}_{a_2,a_1}\right)^2 - 2 \left(\mathcal{F}^{(i)}_{a_1,a_2}\mathcal{F}^{(j)}_{a_2,a_3}\mathcal{F}^{(i)}_{a_3,a_4}\mathcal{F}^{(j)}_{a_4,a_1}\right)\,.
\end{align}
While the first term is the direct contraction of the basis, one would expect that the second one arises with a minus sign due to Grassmann nature of the fermionic fields.
Note also that we have chosen a specific order how to perform the derivatives with respect to the fermion fields.
While these differentiation operation commute in the bosonic case, they do \textit{not} in the case of Grassmann-valued fields.
The order chosen in the present work leads to a projection onto the dressing rather than on its negative.
This leads us to the parametrisation of the four-fermion part of the effective action in terms of \labelcref{eq:CurrentSquared}. It is associated with the basis $\mathcal{B}$, see \labelcref{eq:4FcurrentBasis}, and reads 
\begin{equation}
	\Gamma_\textrm{4f}\left[\Phi\right] = \sum_{i=1}^{N_\textrm{4f}} \lambda_{i,\mathcal B} \, \left(\bar \psi^{a_1} 	\mathcal{F}^{(i)}_{a_1,a_2}\psi^{a_2}\right)\, \left( \bar \psi^{a_3} \mathcal{F}^{(i)}_{a_3,a_4}  \psi^{a_4} \right) \,.
	\label{eq:Gamma4f}
\end{equation}
Complete bases for the momentum-independent four-fermion terms in two and 2+1 flavour theories can be found in \cite{Braun:2019aow, Braun:2020mhk}. In \labelcref{eq:Gamma4f}, we have suppressed the momentum dependence of the four-fermion dressing $\lambda_{i,\mathcal B}$ for the sake of readability.
Furthermore, we have employed the same contraction prescription for operators and field degrees of freedom as defined above.
The scalar product for the four-fermion basis~$\mathcal B$, see \labelcref{eq:scalarProduct} above, leads to the following prescription for the projection onto the dressings~$\lambda_{i,\mathcal B}$: 
\begin{equation}
	\hat P_{\lambda_{i,\mathcal B}}^\alpha = (g_\mathcal{B}^{-1})_{ij}\mathcal{O}^{(j),\alpha}
	\,.
\end{equation}
For a given four-fermion dressing $\lambda_{i,\mathcal B}$, we eventually find
\begin{equation}
	\hat P_{\lambda_{i,\mathcal B}}^\alpha \Gamma_{\textrm{4f},\alpha} = \langle P_{\lambda_{i,\mathcal B}} ,\Gamma_\textrm{4f} \rangle =  \lambda_{i,\mathcal{B}}
	\,. 
\end{equation}
The collective index~$\alpha$ of~$\Gamma_{\textrm{4f},\alpha}$ specifies the order of the derivatives,
\begin{equation}
    \Gamma_{\textrm{4f},\alpha} = \Gamma_{\textrm{4f},a_1a_2a_3a_4} = \frac{\delta}{\delta\psi^{a_1}}\frac{\delta}{\delta\bar\psi^{a_2}} \frac{\delta}{\delta\psi^{a_3}}\frac{\delta}{\delta\bar\psi^{a_4}}\Gamma_\textrm{4f}
    \,.
    \label{eq:G4Derivative}
\end{equation}
In  \labelcref{eq:G4Derivative}, $\alpha$ is given by $\alpha = (a_1,a_2,a_3,a_4)$, where the indices $a_i$ are associated to fermionic and anti-fermionic fields as above.
Finally we remark that the fermionic four-point function can be written in terms of the scalar product defined in \labelcref{eq:scalarProduct} via
\begin{equation}
	\Gamma_{\textrm{4f},\alpha} \simeq (\langle \,\cdot\, , \Gamma_\textrm{4f}\rangle)_\alpha
	\,.
\label{eq:fourFermiVertex}
\end{equation}
Up to now, we have assumed that the underlying basis is known and this has been used implicitly in all projection procedures.
However, the choice of suitable bases, or a respective set of tensors that defines a basis, may be challenging due to the number of degrees of freedom. 
For example, for two quark flavours in QCD (three colours), the basis of four-fermion interactions in the point-like limit has already ten elements, see e.g.~\cite{Mitter:2014wpa, Cyrol:2017ewj,  Braun:2019aow}, while for 2+1 flavours the size of the basis in the point-like limit increases significantly to 26 elements~\cite{Braun:2020mhk}.
Taking into account momentum dependences, i.e. going beyond the point-like approximation of these interactions, the number of elements increases by an order of magnitude or even more. 

This suggests the following procedure for the construction of a basis. 
For the description of the generic principle we stick to the example of QCD: In a first step, an over-complete set of basis elements should be written down by simply considering all tensor combinations of colour, flavour, and Dirac tensors, with the symmetries of QCD as the only constraint. This leaves us with the task to remove redundant elements from the over-complete set of tensors in the second step. 
To begin with, over-completeness is signalled by a singular non-invertible metric. Even on the level of the inner product this property can be observed.
In case of an over-complete basis the sesquilinear form~$\langle \cdot,\cdot \rangle$ may be only positive semi-definite rather than positive definite, which is a necessary condition for a proper basis. Hence, the basis can be reduced until the inner product is positive definite, leaving us with a complete basis with the minimal number of elements.  

In summary, these considerations lead us to a general construction principle for a complete basis for some given interaction:
\begin{enumerate}
	\item Write down a maximal set~$\mathcal{T}_\textrm{max}$ of linearly independent tensors for the interaction, respecting all symmetries of the theory under consideration.
	\item Reduce the set~$\mathcal{T}_\textrm{max}$ to a complete, minimal basis~$\mathcal{T}$ by means of the aforementioned sesquilinear form~$\langle\cdot,\cdot\rangle$ to obtain an inner product space~$\big(\text{span}(\mathcal{T}),\,\langle\cdot,\cdot\rangle\big)$.
\end{enumerate}
Notably, the second step  also takes care of a space reduction due to Fierz-identities and/or the Grassmann nature of specific fields.

Finally, we will briefly explain what we mean with \textit{basis elements} and \textit{vertices} in our terminology. This distinction is relevant for the explicit usage of our Mathematica package \TensorBases described in \Cref{sec:package}.\\[-2ex] 

\textit{Vertices} are the field-derivatives of the effective action noted as $\Gamma_{a_1\alpha_2\dotsc a_n}$. With \labelcref{eq:Falpha} this reads 
\begin{equation}
	\Gamma_{a_1 a_2 \ldots a_n}[\Phi] =
	\frac{\delta}{\delta \Phi^{a_1}} \frac{\delta}{\delta \Phi^{a_2}} \ldots \frac{\delta}{\delta \Phi^{a_n}} \Gamma[\Phi]
	\,.
	\label{eq:Galpha}
\end{equation}
The $\Gamma_{a_1 a_2 \ldots a_n}[\Phi_0]$ are the vertices that enter the diagrams within a vertex expansion about a background $\Phi_0$. In that sense, together with the propagators they are part of the Feynman rules for the derivation of the (one) loop diagrams in the flow equations for (inverse) propagators and vertices. Hence, they may be interpreted as off-shell scatterings between multiple particles. 

For example, a specific vertex, indicated by the superscript $v_n$ in $\Gamma^{(v_n)}_{\alpha}$, can be expanded around $\Phi_0=0$ in terms of tensor \textit{basis elements} $v^{(v_n,i)}_{\alpha}$ with a complete basis ${\cal V}$,  
\begin{equation}
	\Gamma^{(v_n)}_{\alpha} = \sum_{i=1}^{N_v}\,\lambda_{i,\mathcal V}\,v^{(v_n,i)}_{\alpha}
	\,.
	\label{eq:Gammavi} 
\end{equation}
with dressings $\lambda_{i,\mathcal V}$. Here, $\alpha$ is a collective index containing the indices~$\alpha=(a_1 \dots a_{n_v})$. In \labelcref{eq:Gammavi} we restrict ourselves to bases ${\cal V}$, where each basis element 
$v^{(v_n,i)}_{\alpha}$ carries the full symmetry of the theory and the $n$th order derivative operator (crossing symmetry). The associated term $\Gamma^{(v_n)}$ in the effective action can be reconstructed as 
\begin{equation}
 \Gamma^{(v_n)}[\Phi]=	 \frac{1}{\mathcal{N}^{(v_n)}}\,	\Gamma^{(v_n)}_{a_1\cdots a_n} \Phi^{a_1} \cdots\Phi^{a_n} 
	\,. 
\label{eq:vertexReconstruct}
\end{equation}
Furthermore, $\mathcal{N}^{(v_n)}$ is the overall symmetry factor that takes care of the multiplicity of the derivatives. For a theory with one scalar field we have $\mathcal{N}^{(v_n)}=n!$.  Using the expansion \labelcref{eq:Gammavi} for the vertex in \labelcref{eq:vertexReconstruct}, we are led to  
\begin{align}
	\Gamma^{(v_n)}[\Phi] = \frac{1}{\mathcal{N}^{(v_n)}}\,\sum_{i=1}^{N_v}\,\lambda_{i,\mathcal V}\,v^{(v_n,i)}_{\alpha}\,\Phi^{a_1}\dotsc\Phi^{a_{n_v}}
	\,.
\label{eq:vertexExpansion}
\end{align}
Note that the construction with \labelcref{eq:vertexReconstruct} can also be taken as a recipe to obtain explicit basis elements which carry the full symmetries of the theory.

In many cases, one may not start with a fully symmetric vertex basis as in \labelcref{eq:Gammavi}, but chooses a more easily derived basis $\mathcal{B}$. This basis may for example not realise the crossing symmetries of the theory.
We can again expand the vertex action in terms of $\mathcal{B}$ as
\begin{equation}\label{eq:basisExpansion}
	\Gamma^{(v_n)}[\Phi] = \sum_{i=1}^{N_v}\,\lambda_{i,\mathcal B}\,b^{(v_n,i)}_{\alpha}\,\Phi^{a_1}\dotsc\Phi^{a_{n_v}}
	\,.
\end{equation}
To obtain the corresponding~$v_\alpha^{(v_n,i)}$, i.e. the basis elements of the vertex basis~$\mathcal V$ connected to $\mathcal B$ on the level of the effective action, one can perform a simple matrix transformation.
Using the inner product defined by \labelcref{eq:scalarProduct}, we can give the map between vertices and basis elements as
\begin{equation}
	v^{(i)}_\alpha = S^{(b-v)}_{\alpha\beta}b^{(i)}_\beta
	\,,
\end{equation}
where the transfer matrix~$S^{(b-v)}$ is the matrix associated with the inner product given by
\begin{equation}
  \langle a, b\rangle = a^\textrm{T}\cdot S^{(b-v)}\cdot b
  \,,\qquad\forall\,a,b\in V
  \,,
\end{equation}
where $V$ is the vector space of all tensors belonging to $v_n$.
In the example above, this leads directly to \labelcref{eq:fourFermiVertex}, i.e. the four-fermion vertex. 
Note that, for a given set of vertex basis elements~$\{v^{(i)}\} \subset \mathcal V$, one can again define projectors
\begin{equation}
	\hat{P}_{\lambda_{i,\mathcal B}}^\alpha = (\langle v^{(i)}, v^{(j)}\rangle_\textrm{can})^{-1}
	\,.
\end{equation}
Here,~$\langle\cdot,\cdot\rangle_\textrm{can}$ is the standard canonical inner product on the tensor space given by $\langle a,b\rangle_\textrm{can} = a_\alpha b^\alpha$. 
This inner product should not be confused with the definition of the inner product entering the metric, see  \labelcref{eq:metric}.
With this at hand, we eventually arrive at
\begin{equation}
	\hat P_{\lambda_{i,\mathcal B}}^\alpha F_\alpha[\Phi] =  \lambda_{i,\mathcal{B}}
	\,.
\end{equation}
The definitions of projections as discussed above is useful for certain basis constructions which are done directly within the vertex picture, e.g. the construction of a basis for the three-gluon vertex with optimised symmetry decompositions as done in \cite{Eichmann:2014xya, Eichmann:2015nra, Eichmann:2025etg}.
Although the difference is superficial and given by a simple matrix transformation, the tensor basis~$\mathcal B$ defined on the level of the effective action should not be mixed up with the corresponding vertex basis~$\mathcal V$ defined on the level of $n$-point functions in explicit applications.

We close this discussion with the remark that the basis elements $v^{(v_n,i)}_{\alpha}, b^{(v_n,i)}_{\alpha}$ are assumed to be cutoff-independent and hence the dressings $\lambda_{i,\mathcal B}$ carry the full cutoff dependence of the vertex. The respective RG-invariant couplings are composed from products and ratios of the dressings, for a comprehensive discussion see for example~\cite{Ihssen:2024miv}. 

\section{Restriction to Subspaces and Basis Orthogonalisation}
\label{sec:optimisation}

Employing a full basis may be either not feasible due to the size of the resulting system or unnecessary due to the irrelevance of the contributions of a subset tensor structures. Moreover, the computational effort of solving the full system or already the computer-algebraic step of deriving the full functional equations may exceed the capacities of the computers at hand. In these cases, a systematic reduction of the size of a given system and the respective optimisation are chiefly important. In short, one always aims for a reduction of the size of a given system, and thus the computational effort while only minimally affecting the full dynamics.

At the root of these tasks is the access to the relevance ordering of the tensor structures in a given basis and a subsequent rotation of the basis in order to make the descent from dominant tensor structures to irrelevant ones as steep as possible. Seemingly, the most straightforward and easily accessible relevance ordering of a given set of tensor structures is simply to check the relative size of their dressings or couplings. However, this does not do the task justice at all. To begin with, the tensors might be normalised differently and in general the tensors carry different momentum dimensions anyway, which complicates the comparison. Furthermore, the diagrams in a given set of functional relations depend on products of vertices and propagators and the contribution of a given diagram is given by the combinatorial factor of the contraction of all indices as well as the product of all dressings (vertices and propagators), integrated over the loop momentum. Finally, these systems are highly non-linear and even the relative size of diagrams may be deceiving. Loosely speaking, small couplings may have a large impact due to their back-reaction and, vice versa, incidentally large couplings may almost decouple from all observables. Respective analyses in the four-quark system in QCD can be found in \cite{Mitter:2014wpa, Cyrol:2017ewj, Braun:2019aow, Ihssen:2024miv, Fu:2025hcm}.  
 
 In conclusion, a systematic and decisive analysis of the relative importance of different tensor structures must be carried out on the level of observables: 
 
 \textit{If the restriction of the basis for some interaction onto a subspace yields only minimal changes to all relevant observables as well as all correlation functions considered, we call the neglected tensor structures irrelevant.} 
 
Moreover, strictly speaking, this irrelevance only holds true for the observables and correlation functions investigated. For example, it has been shown that the four-quark sector in vacuum QCD is only driven by the scalar-pseudoscalar tensor structure~\cite{Mitter:2014wpa, Cyrol:2017ewj}. Dropping all other tensor structures on the right-hand side of the flow equations (in the loops) does not have a sizeable impact even on the results for the dressings of all tensor structures, leaving aside the value of the chiral condensate and the masses and decay constant of scalar-pseudoscalar mesons. Accordingly, these tensor structures are irrelevant as defined above. However, they are important for the physics of vector mesons, baryons and other hadrons and cannot be neglected within a study of these observables. For a specific, physically relevant, example with the vector meson channels see \cite{Rennecke:2015eba}. Still, their feed back into the flow of the scalar-pseudoscalar dressing is small. This is but one aspects of the \LEGO-principle put forward in \cite{Ihssen:2024miv}.  
 
A further ingredient is the optimisation of the basis at hand as mentioned in the beginning of this section. One aspect of this optimisation is control of the overlap of interaction channels. We illuminate this point at the example of Yang-Mills theories and QCD in the Landau gauge. In these cases, the longitudinal sector decouples from the transversal one and the latter carries the physical information. The metric decays into two blocks, with no overlap between purely transversal and longitudinal basis elements. Furthermore, due to the structure of the propagator, the flow equations for the transverse sector also do not contain any longitudinal information. 
It goes without saying that this is a very special situation, but it remains an open question to which extent this applies to other systems. 

As will be also discussed in the next subsections, a (partial) orthogonalisation of the metric does not imply the decoupling of different basis elements within the fRG approach as well as other functional approaches. 
We distinguish two conceptually different but related mixings:
\begin{enumerate}[(i)]
	\item Mixing due to the projection
	\item Mixing in the flow equations
\end{enumerate} 
Evidently, basis optimisation is key to converging to a minimal and stable systematic expansion scheme, and we shall discuss it and its limitations in the following two subsections. Optimisation of (i), i.e. the reduction of the mixing due to the projections, is discussed in \Cref{sec:MetricOptimisation}. This concerns optimisation procedures on the level of the metric, and loosely speaking corresponds to an optimisation of the left-hand side of the flow equation. Optimisation of (ii), i.e. the reduction of the dynamical mixing of the flow diagrams, is discussed in \Cref{sec:FlowOptimisation}. This can be considered as an optimisation via maximal disentanglement on the right-hand side of the flow equation. 

\subsection{Metric optimisation}
\label{sec:MetricOptimisation}

An apparent optimisation of a given tensor basis with a metric $g_{ij}$ is provided by a diagonalisation. This amounts to finding a transformation or basis rotation ${\cal R}$ with 
\begin{align} 
g_{ij}\stackrel{{\cal R}}{\longrightarrow} g'_{ij}\,,
\end{align}
which renders the new metric $g'_{ij}$ as diagonal as possible. This property is then transferred to the inverse metric as well. The standard Gram-Schmidt orthogonalisation procedure is a simple way to always achieve a fully orthogonal basis. Moreover, it can be readily applied to any basis discussed in the present work. One selects one basis element as the ``anchor" of the transformation. Then, assuming that already a subset of $n$ basis elements have been orthogonalised, one adjusts the next basis element to be orthogonal to this subset, hence enlarging the size of the subset to $n+1$. Iteration eventually leads to a fully diagonal metric and hence an orthogonal basis.
We emphasise that the resulting basis depends on the numbering of the original basis with tensor elements ${\cal O}^{(i)}$ with $i=1,...,N_\textrm{max}$ where $N_\textrm{max}$ is the size of the basis. Accordingly, 
the orthogonalisation procedure is not unique but differs for all permutations of $i$, e.g. for $i=1,2,...,N_\textrm{max} \to i=2,1,....,N_\textrm{max}$. 
Furthermore, the first element is the only one which is certainly left unchanged by the orthogonalisation which may be even desired if the channel has a special meaning. In the general case, one may consider all permutations for maximising the steepest descent in relevance. A relevant example for the physics-informed selection of the first basis element is again provided by four-fermion interactions in vacuum QCD. There, we know that the dynamics is almost solely carried by the scalar-pseudoscalar channel~$\mathcal L_{(\sigma-\pi)}$. This channel carries the dynamics of the spontaneous (strong) breaking of chiral symmetry. Moreover, it carries the pion resonances. Hence, it governs the dynamics of QCD in the deep infrared where chiral perturbation theory is an exceedingly well-working effective theory. Accordingly, this channel should be chosen as the first basis element. We hasten to add that the Fierz ambiguity also entails that the definition of the scalar-pseudoscalar channel is not unique and a full optimisation has to test different definitions of this channel as well. Note also that similar insights into relevance-ordering of interactions as well as momentum channels are also available, e.g. for the 2d Hubbard model in condensed-matter applications~\cite{Metzner_2005, Denz:2019ogb} and further condensed matter systems, see, e.g.~ \cite{PhysRevB.79.085116, PhysRevB.108.235120, Kugler:2017ipn, 10.1143/PTP.112.943} or the reviews \cite{Dupuis:2020fhh, Platt_2013, Metzner:2011cw}.  
When working in a Fierz-complete setting however, the actual choice of the basis is in principle irrelevant. No information gets lost, even though a phenomenological interpretation of the basis elements themselves might be difficult.

This picture dramatically changes when working in a truncation of the full basis. To be more concrete, consider the scalar-pseudoscalar channel~$\mathcal L_{(\sigma-\pi)}$ to be in our basis. Let us choose $\mathcal L_{(\sigma-\pi)}^{\mathcal R}$ to be in the basis as well.
This latter channel is slightly rotated but still almost parallel to the first one (in the sense that $\langle\mathcal L_{(\sigma-\pi)},\mathcal L_{(\sigma-\pi)}^{\mathcal R}\rangle \gg 0$).
Since we have two nearly parallel channels in our basis, at least one of the eigenvalues of the resulting metric becomes close to zero.
Consequently, this leads to a singularity in the inverse metric, which eventually enters the projection.
For example, projecting the fish diagram, see \labelcref{eq:Bonbon}, onto this basis, the flow equation will necessarily mix these two channels. Since they have a large overlap, it is, loosely speaking, not clear ``how much" of the diagram is projected onto the first channel and how much onto the second.
Suppose that, in a second step, we then truncate our basis to a single-channel approximation only containing $\mathcal L_{(\sigma-\pi)}$. After the projection, we will then have lost information as the two channels are not orthogonal to each other. Even worse, we do not even know how much information gets lost in this way.

To conclude, we recommend to first divide a given basis at least in orthogonal sectors if one wants to restrict to a subspace of the full tensor basis. These sectors can be chosen such that they only lightly couple through the flow equations.

Potentially, both a full orthogonalisation and a very singular choice of basis elements (e.g. almost parallel ones) can lead to a worse truncation due to either missing or overly strong mixing effects.
In a Fierz-complete setting, the choice of the basis is of course irrelevant (if one resolves the full momentum dependence). However, for truncated studies, it may still be advantageous and easier interpretable to choose phenomenologically motivated channels.

Finally, we emphasise again that even a full orthogonalisation of the basis does not go hand in hand with a disentanglement of the flow equations.
Nevertheless, there are situations, where an orthogonalisation of the metric implies some simplifications on the level of flow equations. An example would be purely gluonic flows in the Landau gauge. As both the gluon and ghost propagators have essentially no tensorial sub-structure, the orthogonality of the metric transfers at least partially directly to a decoupling of different channels within the flow equations. 
In most cases, however, flow equations mix different couplings due to the different diagrams contributing to the flow. We shall discuss this aspect in the following subsection.

\subsection{Flow optimisation}
\label{sec:FlowOptimisation}

A complementary approach to basis optimisation (i) on the level of the metric, is to perform an optimisation at the level of the flow equation (ii), as discussed in the introduction of \Cref{sec:optimisation}.
This approach takes over the idea of basis optimisation but applies it to the diagrammatic building blocks of the flow equations.

In view of the flow equation, this corresponds to a disentanglement of different basis elements on the right-hand side of the flows.
The guiding principle is then to obtain a maximally decoupled set of flow equations for vertices.

As we shall explain below, a full decoupling of flow equations can be shown to be impossible in the most general case. Therefore, we shall set a less ambitious goal and assume in the following that we are only interested in a single, dominant tensor channel of a given vertex. The aim is then to decouple the flow equations for the different channels in such a way that the channel of interest is affected as little as possible by the other channels.
From this perspective, it is irrelevant whether the channel of interest still feeds into the other channels. Rather, we are looking for a subsystem which is ``as closed as possible", in the sense that other channels do not influence the channel of interest.
The same consideration may be performed for any set of dominant channels. The strategy for optimisation described below is then simply applied to a block of dressing instead of a single one.

Let us make our general strategy towards flow optimisation more concrete at the example of four-fermion flows in an NJL-type model.
We do not restrict ourselves to a specific system, but stick to the most general case of~$N_\textrm{f}$ flavours,~$N_\textrm{c}$ colours and furthermore allow finite temperature and chemical potential. Note that these assumptions are not relevant for the actual optimisation of the flow.

To be more concrete, we restrict ourselves to a purely fermionic formulation of the NJL-type model and only allow for four-quark interactions.
Thus, the only diagrams feeding into the four-fermion interactions are given by fish diagrams
\begin{equation}\label{eq:Bonbon}
	\includegraphics[width=0.14\linewidth]{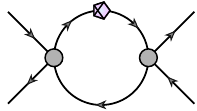}
	\put(-9,15){$\lambda_k$}
	\put(-69,15){$\lambda_j$}
	\put(-122,15){$\lambda_j{M}_{jk}\lambda_k=$}
	\put(5,15){$+\,\dotsc\,.$}
\end{equation}
For~$N_v$ basis elements of the four-quark basis, there are in general~${N_v\choose 2}$ possible fish diagrams contributing to the RG flow.
The four-fermion dressings are denoted by~$\lambda_i$, where we have dropped, for the sake of simplicity, the index~$\mathcal B$, which indicates the chosen basis.
Each contribution on the right-hand side of the flow equation of a given four-quark coupling is of order~$\sim\lambda_i \lambda_j$.
In the diagrammatic equation above,~$\lambda_j$ and~$\lambda_{k}$ are the four-quark dressings, solid lines are fermion propagators, and the grey circles are fully dressed four-quark vertices. Finally, the diamond is a (generalised) regulator insertion.
In general, it is possible to write the different contributions to the flow equation in the form of a matrix multiplication~$\lambda_{j}M_{jk}\lambda_{k}$, where the matrix~$M_{jk}$ contains regularised 1PI loop integral contributions from the fermionic loops.
As discussed in \Cref{sec:inner_product}, however, to obtain the flow equation of a specific four-quark dressing~$\lambda_{i}$, we need to perform a projection of the general flow equation.
To this end, we may now construct the following third-order tensor
\begin{equation}
	D_{ijk} = \hat P_{\lambda_i}^\alpha {M}_{jk,\alpha}
	\,,
\end{equation}
which results from a projection of the ``loop diagram matrix" onto a specific dressing.
For the details of the projection procedure, we refer the reader to \Cref{sec:inner_product}.
In any case, with this third-order tensor at hand, the flow of the coupling~$\lambda_i$ can then be written as 
\begin{equation}
	\partial_t \lambda_i = \frac12\sum_{j,k} D_{ijk}\lambda_j\lambda_k
	\,.
\end{equation}%
Maximal decoupling of the above flow equation is achieved if we manage to diagonalise the full third-order tensor~$D_{ijk}$, i.e.
\begin{equation}
	\hat P_{\lambda_i}^\alpha {M}_{jk,\alpha} = 0\quad\forall\, j\neq i,\,k \neq i
	\,.
\end{equation}
In general, this is not possible.
To show this, we use Theorem 1 and Corollary 1 from \cite{aberth:1967}: A Cartesian tensor $A_{i_1\dotsc i_n}$ can be transformed so that it is diagonal in the indices $i_1,\ldots,i_n$ if and only if it is symmetric in $i_1,\,i_2$ and the tensor
\begin{equation}
  		A_{t i_1\dotsc i_n}A_{t j_2 \dotsc j_n}
  		\,,
\end{equation}
is symmetric in the indices~$i_2$ and~$j_2$. This concludes the theorem. 

In terms of the present situation, the first requirement translates simply to $D_{ijk} = D_{jik}$. The second requirement demands symmetry of
\begin{equation}
\label{eq:DijkDilm}
	D_{ijk}D_{ilm} = \hat P_{\lambda_i}^\alpha {M}_{jk,\alpha} \hat P^\beta_{\lambda_i} {M}_{ml,\beta}
	\,,
\end{equation}
under the exchange~$m \leftrightarrow j$ for all~$j,k,l,m$. Note that summation over~$i$ is tacitly assumed in \labelcref{eq:DijkDilm}. 

For example, restricting our NJL-example from above to only two tensor structures, the second requirement of this theorem corresponds to the statement that
\begin{equation}
  	\includegraphics[width=0.14\linewidth]{four_quark_flow_1}
  	\put(-101,15){$\sum_i P_{\lambda_i}\,\lambda_1$}
  	\hspace{30pt}
  	\includegraphics[width=0.14\linewidth]{four_quark_flow_1}
  	\put(-105,15){$\lambda_2\,\cdot P_{\lambda_i}\,\lambda_1$}
  	\hspace{50pt}
  	\includegraphics[width=0.14\linewidth]{four_quark_flow_1}
  	\put(-126,15){$\lambda_2\,=\,\sum_i P_{\lambda_i}\,\lambda_1$}
  	\hspace{25pt}
  	\includegraphics[width=0.14\linewidth]{four_quark_flow_1}
  	\put(-103,15){$\lambda_1\,\cdot P_{\lambda_i}\,\lambda_2$}
  	\put(-10,15){$\lambda_2$}
\end{equation}\\
In general, there exists no reason why such an identity should exist on the level of diagrams. Even worse, $D_{ijk} = D_{jik}$ assumes a symmetry between the projection operators and the flow equations.

The above construction straightforwardly generalises to flows where the vertex appears more than twice. If the relevant vertex of the interaction occurs within the flows at most~$i$ times, the above construction can be performed, but this leads to the full diagonalisation of a tensor of rank~$i$. 

\begin{figure}[t]
	\includegraphics[width=0.14\linewidth]{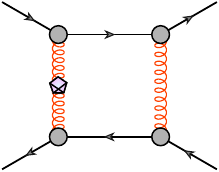}
	\put(-12,37){$g_s$}
	\put(-12,13){$g_s$}
	\put(-63,13){$g_s$}
	\put(-63,37){$g_s$}
	\hfil
	\includegraphics[width=0.14\linewidth]{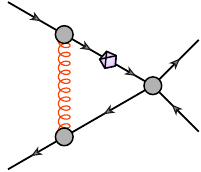}
	\put(-10,25){$\lambda_k$}
	\put(-60,13){$g_s$}
	\put(-60,40){$g_s$}
	\hfil
	\raisebox{3.5mm}{
	\includegraphics[width=0.14\linewidth]{four_quark_flow_1}
	\put(-9,15){$\lambda_k$}
	\put(-69,15){$\lambda_j$}
	}
	\caption{1PI diagrams contributing the the RG flow of the four-fermion interaction in QCD. Diagrams with the same topology but with the regulator insertion attached to other internal lines are not shown. The box diagram~(left) is proportional to the fourth power $g_s^4$ of the strong coupling $g_s$.  The triangle diagrams (middle) are proportional to products of $g_s^2$ and the four-fermion dressings, $\lambda_k g_s^2$ , and the fish diagrams (right) are proportional to $\sim \lambda_{j}\lambda_{k}$.}
	\label{fig:TriangeAndBox}
\end{figure}

Thus, a full diagonalisation is in general not possible, even at the level of a NJL-type model, where only fermionic degrees of freedom are taken into account.
Turning our focus to QCD and its fundamental degrees of freedom, namely quarks and gluons, the situation may become even more involved.
The diagonalisation of~$D$ becomes more complicated when allowing for gluonic degrees of freedom contributing to the flow of four-quark interactions, see the diagrams in \Cref{fig:TriangeAndBox}, since additional diagrams have to be taken into account.

Nevertheless, let us present a strategy to perform a flow optimisation on a very pragmatic level. 
To this end, it is instructive to consider the dynamics of four-quark RG flows in QCD.
There, we know that the two-gluon exchange diagram (box diagram) is dominating the flow of the four-fermion couplings at large RG scales, see \Cref{fig:TriangeAndBox} (left).
This has been shown in many QCD studies \cite{Gies:2005as, Braun:2006jd, Mitter:2014wpa, Cyrol:2017ewj, Braun:2019aow} and is explicitly investigated in~\cite{Ihssen:2024miv}.
At intermediate RG scales, the triangle diagram, see \Cref{fig:TriangeAndBox} (middle), becomes more relevant until the fish diagram eventually dominates the running of the four-fermion interaction.

Keeping the situation in QCD in mind, let us now assume again that we are interested in a single four-fermion channel~$\lambda_i$.
The dressing of interest should be chosen in such a way that it is dominantly driven at large RG scales, i.e. by the box diagram.
In other words, since all four-fermion dressing are initially set to zero in QCD flows, we choose the specific dressing of interest as the one that is mainly fed by the box diagram.
Of course, the running of the strong coupling constant is dominantly driven by the Yang-Mills sector and less effected by the four-quark interaction.
Therefore, there is only a sub-leading back coupling from the four-fermion interaction into the flow of the strong coupling.
With this at hand, let us again consider the running of the four-quark dressing driven by the fish diagrams
\begin{equation}
	\partial_t \lambda_i = \frac12\sum_{j,k} D_{ijk}\lambda_j\lambda_k
	\,.
\end{equation}
Starting from that, we construct a new basis, which is orthogonal to the the basis element~$j\neq i$ associated with the the running of the coupling~$\lambda_i$.
This can be done by employing the Gram-Schmidt orthogonalisation procedure to diagonalise the matrix~$D_{ijk}$ with respect to the indices~$j$ and~$k$.
This yields
\begin{equation}
	\partial_t \lambda_i = \frac12\sum_{j,k} \tilde D_{ijk} \tilde\lambda_j\tilde\lambda_k \quad \text{with} \quad \lambda_i = \tilde\lambda_i \quad \text{and} \quad \tilde D_{iii} = D_{iii}
	\,,
\end{equation}
where the new matrix~$\tilde{D}_{ijk}$ fulfils the property~$\tilde{D}_{ijk} \sim \delta_{jk}$.
Note that we have performed the following basis change: $\tilde\lambda_i = S_{in}\lambda_{n}$ with~$\tilde\lambda_{i} = \lambda_{i}$ and~$\tilde D_{ijk} = D_{ilm} S^{-1}_{lj}S^{-1}_{mk}$ with~$\tilde D_{iii} = D_{iii}$.
Since the couplings~$\lambda_j$ depend on the RG scale, also the transition matrix~$S$ does.
However, this dependence may be absorbed into the matrix~$\tilde D_{ijk}$ and the new dressings~$\tilde\lambda_{i}$.
Eventually, we arrive at a flow equation for~$\lambda_i$ which, due to the partial diagonalisation, only contains contributions~$\sim \lambda_j^2$.
\begin{equation}
	\partial_t \lambda_i = \frac12\sum_{j} \tilde D_{ijj} \tilde\lambda_j^2
	\,.
\end{equation}
Since the coupling~$\lambda_{i}$ is dominant  at large RG scales by construction, the term~$\lambda_i^2$ is mainly relevant. Contributions of order~$\lambda_j^2$ with~$j \neq i$ are suppressed.
In particular, these contributions are more suppressed than terms~$\lambda_i\lambda_j$, which are eliminated by the new basis.

In addition to the pragmatic procedure described above, a partial optimisation can be obtained by performing an approximate diagonalisation of the whole third-order tensor~$D_{ijk}$.
Within the context of signal processing, the approximate diagonalisation of rank three and higher order tensors has been studied, e.g. in \cite{li20193rdOrderTensors1}. However, such a treatment requires a iterative numerical evaluation of the full flow structure of the corresponding tensor basis, together with an implementation of an optimisation algorithm. This presents a sizeable numerical challenge.
In addition to that, the whole basis is being transformed in such an approach. 

In conclusion, we have set up an optimisation approach for maximising the physics of a given approximation of functional approaches, while minimising the computational tasks and the systematic error. It can be readily implemented and has many applications. In short, it shows great potential.

\section{The momentum structure of tensor bases}
\label{sec:momenta}

In this section, we will discuss the momentum dependence and parametrisation of tensor bases, projections and associated dressings. 
This is a crucial aspect of basis optimisation, as certain choices of momentum-dependent bases may lead to irregularities in the metric and thus also in the associated projection operators. This may lead either to ill-conditioned numerical problems which artificially require high numerical precision, or even worse ill-defined projections at specific momentum configurations.

To investigate this aspect of basis optimisation, we will first introduce some convenient momentum parametrisations for three-point functions in \Cref{sec:momParam}. 
Then, in the second subsection, we investigate explicitly the quark-gluon vertex and the choice of the momentum structure within the basis as an important example where one can encounter momentum irregularities that one needs to remove by pertinent basis transformations.
Although we keep the discussion in this section on the level of three-point functions, our considerations can be straightforwardly extended to other orders.

\subsection{Momentum parametrisation}
\label{sec:momParam}

For the convenience of the reader, let us first recount parts of \cite{Eichmann:2014xya} which are relevant for our present work. 
We discuss most useful momentum parametrisations for three-point functions. 
For four-point functions, we refer the reader also to \cite{Eichmann:2015nra}.

In general, any dressing belonging to a three-point vertex can be written as
\begin{equation}
	\lambda(p_1,p_2,p_3)
	\,.
\end{equation}
We take the first momentum~$p_1$ to be the bosonic momentum if~$\lambda$ describes a one-boson-two-fermion interaction. It is useful to introduce the vectors
\begin{equation}
	Q = p_1\,, \qquad
	k = \frac{p_3-p_2}{2}
	\,.
\end{equation}
In order to find a better representation, one can first switch to the related Lorentz invariants
\begin{equation}
	t = \frac{Q^2}{4}
	\,,\qquad
	\xi = \frac{4 k^2}{3 Q^2}
	\,,\qquad
	z = \frac{k \cdot Q}{\norm{k} \norm{Q}}
	\,,
\end{equation}
with $t\geq0$,~$\xi\geq0$ and~$z\in [-1,1]$.
We can now conveniently introduce the variables
\begin{align}
	\mathcal{S}_0^2 &= 2\,t (1+\xi) = \frac{1}{3}\left( p_1^2 + p_2^2 + p_3^2 \right)
	\,,\\[1ex]
	a &= \frac{2 z \sqrt{\xi}}{\xi + 1} = \sqrt{3} \frac{p_3^2 - p_2^2}{p_1^2 + p_2^2 + p_3^2} 
	\,,\\[1ex]
	s &= \frac{\xi - 1}{\xi + 1} = \frac{p_2^2 + p_3^2 - 2 p_1^2}{p_1^2 + p_2^2 + p_3^2}
	\,.
\end{align}
While $S_0$ represents the average momentum of all particles, $a$ and $s$ are both angular variables, constrained to the unit disk by $a^2+s^2 \leq 1$.
The~$(a,s)$-unit-disk combined with the average momentum~$\mathcal{S}_0 \in [0, \infty)$ describes the entire phase space as a cylinder. 
The symmetric point is located at the centre of the disk, whereas the soft limits (i.e. when one of the momenta is set to zero) are positioned at the cube roots of 1.
	
In actual implementations, it is useful to choose coordinates more pertinent to the unit disk spanned by~$a$ and~$s$. To this end, we replace~$a$ and~$s$ by
\begin{align}\notag
	\mathcal{S}_1 &= a^2 + s^2 \in [0,1]
	\,,\\[1ex]
	\mathcal{S}_\varphi &= \text{arctan2}(a,s)\in [0,2\pi)
	\,,
\end{align}
where~$\text{arctan2}(a,s) = \text{arg}(a+\text{i} s)$ and~$\text{arctan2}\,:\,\mathbb{R}^2\to[0,2\pi)$ is the inverse tangent function extended to the full plane~$\mathbb{R}^2$.

\subsubsection*{Symmetric parametrisation}

The presence of an additional symmetry between the fields partaking in the interaction may simplify the structure of the phase space even further.
In the case of an interaction between three identical bosons, the basis may be chosen such that it reflects this symmetry. 
It is then not necessary to compute the full momentum disk spanned by $a$ and $s$ as it is partially rendered superfluous by the $S_3$-symmetry of the vertex. 
Each $120^\circ$-slice of the angular disc is a copy of the previous one, and we have the choice to use the $S_3$-symmetric parametrisation
\begin{align}\notag
	\mathcal{S}_1 &= a^2 + s^2 \in [0,1]
	\,,\\[1ex]
	\mathcal{S}_2 &=s(3a^2-s) \in [-1,1]
	\,.
\end{align}
This restricts the angle to one slice of~$120^\circ$ of the circle spanned by~$a$ and~$s$.
However, let us make our considerations more concrete by an example.

\subsection{Transverse Quark-Gluon Vertex} 
\label{sec:quarkGluonVertex}

We illustrate the above analysis at the example of the quark-gluon vertex in QCD. Our discussion here draws from respective ones and explicit computations in \cite{Mitter:2014wpa, Williams:2014iea, Williams:2015cvx, Cyrol:2017ewj, Gao:2021wun, Aguilar:2024ciu}

, restricting ourselves here to only the transverse sector.
Generally, we define the channels of the basis as
\begin{equation}
	\mathcal{L}_{A\bar qq}^{(i)} = (2\pi)^d\delta(p+q+r) \, T^a \, \,\Pi^\perp_{\mu\nu}(p)[\mathcal{T}_{A\bar{q}q}^{(i)}]_\nu(p,q,r)\,\lambda_{A\bar qq,i}(p,q)
	\,,
\end{equation}
where we have already implemented momentum conservation and the common element~$T^a$ of the fundamental representation of the colour group and furthermore introduced the transverse projector
\begin{equation}
	\Pi_{\mu\nu}(p) = \delta_{\mu\nu} - \frac{p_\mu p_\nu}{p^2}
	\,.
\end{equation}
We fully suppress the flavour indices of the quarks as the group structure thereof is trivial (i.e. proportional to the one element in flavour space) in the quark-gluon vertex.
The scalar product as introduced in \labelcref{eq:scalarProduct} is rather simple as all particles are distinct,
\begin{equation}
	\langle\mathcal{T}^{(i)},\mathcal{T}^{(j)}\rangle = [\mathcal{T}_{A\bar{q}q}^{(i)}]_{a_1a_2a_3}[\mathcal{T}_{A\bar{q}q}^{(j)}]_{a_1a_3a_2}
    \,.
\end{equation}
Here, we use collective indices where $a_1$ corresponds to the gluon and $a_2$ and $a_3$ to the quarks.	
A possible choice of tensor structures entering the various bases, which are directly constructed from $\slashed D^n$, is consequently given by
\begin{alignat}{2}
		[\mathcal{T}_{A\bar{q}q}^{(1)}]^a_\nu(p,q,r) & =\,i\,\gamma_\nu
		\,,\notag                                                                                              \\[1ex]
		[\mathcal{T}_{A\bar{q}q}^{(2)}]^a_\nu(p,q,r) & =\,(q-r)_\nu
		\,,\notag                                                                                              \\[1ex]
		[\mathcal{T}_{A\bar{q}q}^{(3)}]^a_\nu(p,q,r) &=\,(\slashed q-\slashed r)\gamma_\nu
		\,,\notag                                                                                              \\[1ex]
		[\mathcal{T}_{A\bar{q}q}^{(4)}]^a_\nu(p,q,r) &=\,(\slashed q+\slashed r)\gamma_\nu
		\,,\notag                                                                                              \\[1ex]
		[\mathcal{T}_{A\bar{q}q}^{(5)}]^a_\nu(p,q,r) & =\,(\slashed q + \slashed r)\,i (q-r)_\nu
		\,,\notag                                                                                              \\[1ex]
		[\mathcal{T}_{A\bar{q}q}^{(6)}]^a_\nu(p,q,r) & =\,(\slashed q - \slashed r)\,i (q-r)_\nu
		\,,\notag                                                                                              \\[1ex]
		[\mathcal{T}_{A\bar{q}q}^{(7)}]^a_\nu(p,q,r) & =\,\frac{\textrm{i}}{2}\,[\slashed q,\slashed r]\,\gamma_\nu
		\,,\notag                                                                                              \\[1ex]
		[\mathcal{T}_{A\bar{q}q}^{(8)}]^a_\nu(p,q,r) & =\,\frac{1}{2}\,[\slashed q,\slashed r]\,(q-r)_\nu
		\,.
\end{alignat}
However, we would like to emphasise that an unpleasant issue comes along with this choice as mentioned the introductions of the present section.
To be precise, the first tensor structure~$\mathcal{T}_{A\bar{q}q}^{(1)}$ has a finite overlap with~$\mathcal{T}_{A\bar{q}q}^{(6)}$ which leads to projection singularities. 
In particular, looking at the projector onto the first basis element defined via~$P^{\lambda_{A\bar qq,1}}=g_{ij}^{-1} \mathcal{T}^{(j)}$, the singularities become apparent immediately.
The projector is given by
\begin{equation}
	P^{\lambda_{A\bar qq,1}} = -\frac{1}{64} \mathcal{T}^{(1)} +
		\frac{S_1 \cos(S_\varphi)}{64 \sqrt{3}\, S_0^2 \left(S_1^2-1\right)} \mathcal{T}^{(5)} + 
		\frac{S_1 \sin(S_\varphi)-1}{192\,S_0^2 \left(S_1^2-1\right)} \mathcal{T}^{(6)}
	\,.
\end{equation}
Due to the matrix inversion required for definition of the projectors, also $\mathcal{T}_{A\bar{q}q}^{(5)}$ gets mixed up in the projection onto the first tensor element, although $\langle\mathcal{T}^{(1)},\mathcal{T}^{(5)}\rangle = 0$.
Here, we have already utilised the momentum parametrisation put forward in \Cref{sec:momParam}, which is most useful in the following discussion.
In any case, the projection above is singular for all soft limits which lie at $S_1 = 1$.
However, a redefinition of $\mathcal{T}^{(6)}$ allows us to ``clean up" the first tensor structure.

We would like to stress that this singularity is particularly delicate as the classical tensor structure~$\mathcal{T}_{A\bar{q}q}^{(1)}$ is of utmost importance as it turns out to be the most dominant one.
Additionally, in fRG flow equations, the classical tensor structure is the only quark-gluon vertex already present at the initial scale~$\Lambda$ and it enters the perturbative Slavnov-Taylor identities (STIs), which are relevant for setting appropriate initial conditions.

Of course, the singular mixing is exceptionally problematic if diagrams involve any kind of soft limit. 
This is the case for the infrared limits of both the gluon and the quark propagators.
As the projection is ill-defined on the soft limits, exhibiting poles, even calculating close to these singularities means solving an ill-conditioned problem. 
For actually simple calculation, the numerical accuracy that is required might increase in such cases considerably, just as a consequence of choosing a basis with inherent momentum singularities~\cite{SattlerHardQCD}.
Note also that the above projection with its singularity will affect the RG running of the first tensor structure. This has the potential of destabilising the system and at least introduces a large systematic error. 
A specific example is the matter dynamics of vacuum QCD at low energies: it is governed by dynamical chiral symmetry breaking and the low-energy dynamics is primarily determined by the light quarks and the lightest mesons. 
Thus, the accurate determination of the low-energy part of the quark propagator is crucial and a large systematic error induced by singularities in the projections can spoil the predictive power of the calculation.

A simple disentangling of the sixth tensor structure from the first, by means of the Gram-Schmidt algorithm, leads to much more efficient numerical computations and therefore requires much less precision for excellent results~\cite{SattlerHardQCD}.
In practice, this can be straightforwardly achieved by applying the following redefinition
\begin{equation}
	[\mathcal{T}_{A\bar{q}q}^{(6)}]_\nu(p,q,r) \to\,(\slashed q - \slashed r)\,\textrm{i}(q-r)_\nu - \frac{\langle\mathcal{T}^{(1)},\mathcal{T}_{A\bar{q}q}^{(6)}\rangle}{\langle\mathcal{T}_{A\bar{q}q}^{(1)},\mathcal{T}_{A\bar{q}q}^{(1)}\rangle}[\mathcal{T}_{A\bar{q}q}^{(1)}]_\nu
	\,,
\end{equation}
where
\begin{equation}
	\frac{\langle\mathcal{T}^{(1)},\mathcal{T}_{A\bar{q}q}^{(6)}\rangle}{\langle\mathcal{T}_{A\bar{q}q}^{(1)},\mathcal{T}_{A\bar{q}q}^{(1)}\rangle}	= \frac{S_1 \sin(S_\varphi)-1}{S_0^2 \left(S_1^2-1\right)}
	\,.
\end{equation}
In QCD, however, the fourth and seventh tensor structures are especially important as they affect low-energy observables most, see \cite{Mitter:2014wpa,Cyrol:2017ewj,Gao:2021wun}. Restricting ourselves to the tensors 1, 4 and 7, the associated inverse metric reads
\begin{equation}
g_{ij}^{-1} = \begin{pmatrix}
		-\frac{1}{96} & 0 & 0 \\
		0 & \frac{1}{96 S_0^2 (S_1 \sin(S_\varphi)-1)} & 0 \\
		0 & 0 & \frac{1}{24 S_0^4 \left(S_1^2-1\right)} \\
	\end{pmatrix}
	\,.
\end{equation}
Once again, the seventh tensor structure has a diverging projection in all soft limits whereas the fourth tensor structure diverges in the soft-gluon limit, $S_\varphi=\frac\pi2$. 
In this case, a definite solution is difficult to find. For numerical applications, however, it may be advantageous to shift the diverging terms into the vertex itself to prevent the projection from diverging. 
In practice, this must be decided on a case-by-case basis. 
Of course, the problem is absent, if a symmetric-point approximation $S_1 = 0$ is chosen. Then, the above metric does not carry any angular dependence and thus no angular singularities are present.
In addition to the angular dependences, it may also be useful to consider dimensionless dressings and absorb powers of the average momentum $S_0^n$ into the tensor basis itself, which removes the singularity at $S_0 = 0$ within the projectors.

Lastly, we mention that, while the suggestions given above certainly improve the calculation of the quark-gluon vertex, other parts may still be singular. 
To be more explicit, the flow equations themselves may have removable singularities in the angular variables which can impede the calculation, but are very hard to remove entirely.

\section{Four-Fermion Vertices and Fierz identities}
\label{sec:fourFermiVertex}

In the following, we shall focus on the four-fermion vertex in the point-like limit in more detail.
As already discussed in \Cref{sec:inner_product}, the construction of a basis can be done by writing down all elements which are in accordance with the symmetries of the theory under consideration. As a next step, one must reduce this set until the resulting metric is non-singular, i.e. a maximal linearly independent set of basis vectors has been found. 
The projection procedure on any of the associated dressings can be defined as introduced in \Cref{sec:inner_product}.

For QCD with two massless quark flavours at finite temperature and quark chemical potential, one finds that the basis of the four-fermion vertex consists of ten elements, if we restrict ourselves to the point-like limit, in agreement with~\cite{Braun:2019aow, Braun:2020mhk}. Note that the full basis for scalar four-quark interactions, which includes also momentum-dependent tensor structures, counts 256 tensors \cite{Eichmann:2015cra}. For applications to low-energy physics, the momentum-dependent tensor structures should be less relevant and thus we only focus on the point-like limit.

A phenomenologically motivated choice of basis elements is given by \cite{Braun:2018bik}
\begin{align}\label{eq:basis}
	\mathcal{L}_{(V+A)_\parallel} &= (\bar\psi \gamma_0 \psi)^2 + (\bar\psi i \gamma_0 \gamma_5 \psi)^2
	\notag\\[1ex]
	\mathcal{L}_{(V+A)_\perp} &= (\bar\psi \gamma_i \psi)^2 + (\bar\psi i \gamma_i \gamma_5 \psi)^2
	\notag\\[1ex]
	\mathcal{L}_{(V-A)_\parallel} &= (\bar\psi \gamma_0 \psi)^2 - (\bar\psi i \gamma_0 \gamma_5 \psi)^2
	\notag\\[1ex]
	\mathcal{L}_{(V-A)_\perp} &= (\bar\psi \gamma_i \psi)^2 - (\bar\psi i \gamma_i \gamma_5 \psi)^2
	\notag\\[1ex]
	\mathcal{L}_{(V+A)_\parallel^\dagger} &= (\bar\psi \gamma_0 T^a \psi)^2 + (\bar\psi i \gamma_0 \gamma_5 T^a \psi)^2
	\notag\\[1ex]
	\mathcal{L}_{{(V-A)}_\perp^\dagger} &= (\bar\psi \gamma_i T^a \psi)^2 -(\bar\psi i \gamma_i \gamma_5 T^a \psi)^2
	\notag\\[1ex]
	\mathcal{L}_{(\sigma-\pi)} &= (\bar\psi \psi)^2 - (\bar\psi \gamma_5 \tau_i \psi)^2
	\notag\\[1ex]
	\mathcal{L}_{(S+P)_-} &= (\bar\psi \psi)^2 - (\bar\psi \gamma_5 \tau_i \psi)^2 +(\bar\psi \gamma_5 \psi)^2 - (\bar\psi \tau_i \psi)^2
	\notag\\[1ex]
	\mathcal{L}_{\mathrm{csc}} &= (\bar\psi \gamma_5 \mathcal{C} \tau_2 i \epsilon_a \bar\psi^T)(\psi^T \mathcal{C}\gamma_5 \tau_2 i \epsilon_a \psi)
	\notag\\[1ex]
	\mathcal{L}_{(S+P)_+^\dagger} &= (\bar \psi T^a \psi)^2 - (\bar\psi \gamma_5 \tau_i T^a \psi)^2 +(\bar\psi \gamma_5 T^a \psi)^2 - (\bar\psi \tau_i T^a \psi)^2 
	\,.
\end{align}
Here, $\gamma_\mu$ are the Dirac matrices, $\epsilon_a$ are the antisymmetric generators of the~SU$(3)$ colour transformations.
Furthermore, $T^a_c$ and~$\tau_i$ are the fundamental representations of the generators of the colour gauge group and the SU$(N_f)$, and $\mathcal{C}$ is the charge conjugation matrix.
Of course, a different basis can be straightforwardly constructed by considering linear combinations of the aforementioned basis elements or by performing Fierz transformations.
This property is referred to as \textit{Fierz completeness}.

Furthermore, with the exception of $\mathcal{L}_\mathrm{csc}$, we have written down only basis elements which consist of linear combinations of coupled two-quark currents, i.e. $(\bar\psi\mathcal{F}^{(i)}\psi)(\bar\psi\mathcal{F}^{(j)}\psi)$ (see also \Cref{sec:inner_product}). 
Any kind of four-fermion interaction can be written either in terms of meson-meson or diquark--anti-diquark four-quark channels. The two representations are connected through a Fierz transformation. For example, the reformulation of $\mathcal{L}_\mathrm{csc}$ in terms of meson-meson channels is demonstrated in \Cref{app:fierzIdentitieCSC}.

With respect to the basis \labelcref{eq:basis}, we also note that charge conjugation~$\mathcal C$ is not preserved at finite chemical quark potential due to the quark--anti-quark imbalance. This is not reflected in the above basis. As all currents form the Dirac basis \labelcref{eq:DiracBasis} are either $\mathcal C$-symmetric or $\mathcal C$-antisymmetric, any $\mathcal C$-breaking four-quark interaction requires a combination of two different quark currents with a different $\mathcal C$-parity.
For example, $\bar{\psi}^C\gamma_\mu \psi^C = -\bar{\psi}\gamma_\mu \psi$ and thus the combination $(\bar\psi \psi \, )(\bar\psi\gamma_0\psi)$ would constitute a viable $\mathcal C$-breaking interaction.
In fact, this is the only viable $\mathcal C$-breaking four-fermion interaction.

In general, constructing all (pseudo-)scalar $\mathcal C$-antisymmetric four-quark combinations in the point-like limit (i.e. without operators $\partial_\mu$), we obtain only three possible combinations:
\begin{equation}
	(\bar{\psi}\gamma_\mu\psi)(\bar{\psi}\gamma_\mu\gamma_5\psi)
	\,,\qquad
	(\bar{\psi}\gamma_0\psi)(\bar{\psi}\psi)
	\,,\qquad
	(\bar{\psi}\gamma_0\psi)(\bar{\psi}\gamma_5\psi)
	\,.
	\label{eq:CBreaking}
\end{equation}
In addition to the $\mathcal C$-symmetry, the first term also breaks parity $\mathcal P$. The second one breaks $\mathcal C$-invariance and the axial U$_A$(1)-symmetry. Last but not least, the third one breaks~U$_A$(1),~$\mathcal P$ and~$\mathcal T$ (time-reversal). 
As the QCD action is $\mathcal P$- and $\mathcal T$-symmetric, neither the first nor the third term will be generated in the RG flow. 
However, the second term can arise in the presence of chiral symmetry breaking and will be generated in an approach to QCD at finite baryon chemical potential $\mu_B > 0$.
Since we restrict ourselves to vanishing current quark masses, i.e. to the chiral limit, where the chiral symmetry is intact, we do not include such a term in our four-quark basis.
In addition, we have explicitly checked that the box diagram, see \Cref{fig:TriangeAndBox} (left), does not generate such a contribution, even at finite~$\mu_B$ in the chiral limit.
If not generated by the box diagram as the fundamental building block of four-quark interactions they will not be generated at all.
However, all four-quark tensor structures that are allowed by the symmetries of the theory under considerations, are generated by the box diagram.
Of course, the situation is very different if the chiral symmetry is broken in some way.  
Tensor structures such as \labelcref{eq:CBreaking} are then naturally generated and the size of the four-fermion basis increases significantly. 

With respect to QCD phenomenology, we would like to highlight two aspects: First, the basis \labelcref{eq:basis} contains the sigma-pion channel $\mathcal{L}_{(\sigma-\pi)}$ which becomes resonant at low temperatures and densities, leading to the formation of a scalar condensate $\langle\bar{\psi}\psi\rangle \neq 0$. 
Thus, to describe chiral symmetry breaking, the channel $\mathcal{L}_{(\sigma-\pi)}$ is crucial and is also by far the most dominant channel in this regime~\cite{Mitter:2014wpa,Braun:2019aow}.
Second, at high densities, the diquark channel~$\mathcal{L}_{\mathrm{csc}}$ describing bound off-shell coloured states, the diquarks, has been argued to be most relevant, see, e.g. \cite{Bailin:1983bm,Buballa:2003qv,Alford:2007xm} for reviews. With the fRG approach, the dynamical generation, enhancement, but also suppression of the aforementioned four-quark interactions has been investigated throughout the QCD phase diagram in \cite{Braun:2018bik,Braun:2019aow}.
In particular, at very high densities, there has been renewed interest in the effect of diquark condensation and the properties of the associated colour-superconducting phases. For some recent calculations using perturbation theory, see~\cite{Braun:2022jme,Geissel:2024nmx}. For corresponding fRG studies, we refer to~\cite{Leonhardt:2019fua,Fukushima:2021ctq,Braun:2021uua,Braun:2022olp}.

\begin{figure}[t]
	\centering
	\resizebox {0.35\textwidth} {0.35\textwidth} {\includegraphics{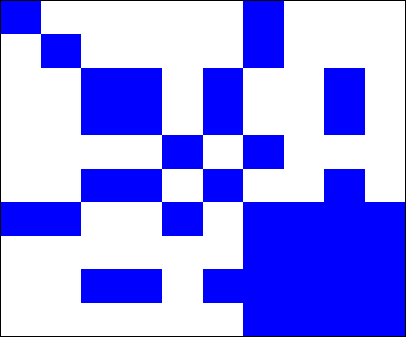}}
	\put(-68,173){$\mathcal{L}_\mathrm{csc}$}
	\put(-37,173){$\mathcal{L}_{\!(\!\sigma\!-\!\pi\!)}$}
	\put(2,55){$\mathcal{L}_\mathrm{csc}$}
	\put(2,22){$\mathcal{L}_{\!(\!\sigma\!-\!\pi\!)}$}
	\caption{Colour-coded sparsity pattern of the metric of \labelcref{eq:basis}. Blue rectangles correspond to a finite overlap, while white fields correspond to a vanishing scalar product. Each rectangle corresponds to a specific combination of two channels. 
	For example, the upper left rectangle corresponds to the scalar product $\langle \mathcal{O}^{(V+A)_\parallel}, \mathcal{O}^{(V+A)_\parallel}\rangle$.
	The other fields are defined accordingly to the ordering of the basis elements \labelcref{eq:basis}. 
	Note that the phenomenologically important colour-superconducting (No. 7) and scalar-pseudoscalar channels (No. 9) (lower right corner) have an overlap with a significant fraction of the channels.}
	\label{fig:metric}
\end{figure}
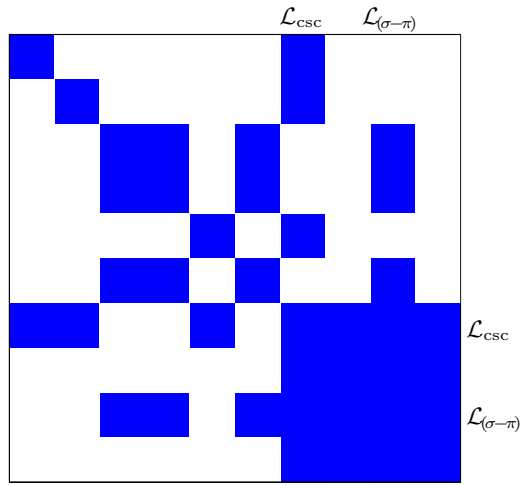

Bearing these observations in mind, it might be useful to look at the metric associated with four-quark basis in~\labelcref{eq:basis} to choose a basis which separates into two blocks, namely the scalar-pseudoscalar together with the colour-superconducting channel and the remaining channels, that are chosen to be in the orthogonal complement of the former.
For a more detailed discussion on this aspect, see \Cref{sec:optimisation}.

Following the procedure outlined in \Cref{sec:inner_product}, let us now calculate the metric of the basis \labelcref{eq:basis}. 
First of all, we observe that the basis element associated with the colour-superconducting channel, namely~$\mathcal{L}_\textrm{csc}$, is not of the form~$(\bar\psi\mathcal{F}^{(i)}\psi\bar)(\psi\mathcal{F}^{(j)}\psi)$.
However, especially for the application of computer algebra systems, it turns out to be useful to rewrite $\mathcal{L}_\mathrm{csc}$ into the common form associated with two coupled mesonic channels.

To this end, we can make use of Fierz transformations to, loosely speaking, "reshuffle" the fermionic fields. In general, there are three different kinds of Fierz transformations, depending on which fermionic fields should be exchanged.
Even though Fierz transformations can be performed for finite momenta, we will restrict ourselves to the point-like limit here.
For a transformation of $\mathcal{L}_\mathrm{csc}$, we need the following Fierz identity,
\begin{equation}\label{eq:FierzIdForcsc}
	(\bar \psi M \bar \psi^T)
	(\psi^T \tilde M \psi)
	=
	\sum_{i,j}
	\textrm{Tr}\left(\mathcal F^{(i)} M \left(\mathcal F^{(j)}\right)^T \tilde M\right)
	\left(\bar \psi \mathcal F^{(i)} \psi\right)
	\left(\bar \psi \mathcal F^{(j)} \psi\right)
	\,.
\end{equation}\\[-2ex]
For the sake of readability, we have dropped the indices.
Furthermore,~$M$ and~$\tilde M$ are in principle arbitrary matrices in colour, flavour, and Dirac space.
The derivation of the Fierz identity \labelcref{eq:FierzIdForcsc} as well as the other Fierz identities mentioned above can be found in \Cref{app:fierzIdentities}.

From \labelcref{eq:FierzIdForcsc}, one can immediately deduce that the colour-superconducting channel is transformed into a form that can be easily used to calculate the metric \labelcref{eq:metric}.
Since the resulting expression is lengthy, the explicit form of the Fierz transformation for the colour-superconducting channel is also only given in \labelcref{eq:cscFierzTrafo}.

Performing the Fierz transformation above allows the metric to be calculated directly in a fully basis-independent manner.\footnote{Of course, the computation can be also performed directly in terms of the original channel~$\mathcal{L}_\mathrm{csc}$. However, using common computer algebra systems, such as Mathematica, this requires a representation-independent implementation of the $\mathcal C$-operator.}
We illustrate the sparsity pattern of the resulting metric in \Cref{fig:metric}. Rows/columns seven and nine correspond to the phenomenologically relevant basis elements~$\mathcal L _{(\sigma-\pi)}$  and~$\mathcal L _{\mathrm{csc}}$. 
Apparently, this metric is sparse. However, this does not imply that the flow equations of the four-fermion dressing functions decouple in the same way. In fact, with the notation as introduced in \Cref{sec:optimisation}, the coupling tensor $D_{ijk}$ for this basis of four-fermion channels is dense, with only very few vanishing elements. This again emphasises the fact that the simplicity of the metric does not generally entail simplifications in the set of flow equations.

\section{The \TensorBases Mathematica package}
\label{sec:package}

The considerations in \Cref{sec:inner_product} gives a very straightforward prescription how to deal with any given tensor basis or vertex basis. Clearly, obtaining the projection operators for a given basis and scalar product is algorithmically simple and, for convenience, we provide a Mathematica package, \TensorBases, to automate these tasks efficiently. 
Furthermore, we provide functions to define new bases, construct bases from sets of basis tensors and reduce such sets with respect to a given scalar product.

\href{https://github.com/satfra/TensorBases.git}{\TensorBases} is open-source and freely available on GitHub and already contains pre-defined and compiled bases for the four-quark vertex in the vacuum limit as well as for the case of finite temperature and/or chemical potential, the quark gluon vertex, and three- and four-gluon vertices. 
It is built on the \texttt{FormTracer} Mathematica package \cite{Cyrol:2016zqb}, which provides very convenient routines to trace tensor structures in loop expressions using the high-performance computer algebra software \texttt{FORM} \cite{Ruijl:2017dtg}.
To install the package in Mathematica, one can simply use the command

{
\vspace{1ex}
\footnotesize
\begin{mmaCell}{Input}
\mmaDef{Import}["https://raw.githubusercontent.com/satfra/main/TensorBasesInstaller.m"]
\end{mmaCell}
}

\noindent
The package is easy to use and comes with a built-in documentation. In particular, using the command \mmaInlineCell{Code}{TBInfo[...]}, one can directly access a detailed documentation of all \TensorBases commands, its functionality, and known bases.
To see all pre-defined bases, simply call

{
\vspace{1ex}
\footnotesize
\begin{mmaCell}{Input}
\mmaDef{TBInfo}[]
\end{mmaCell}
}

\noindent
Note that \TensorBases defines a handful of extensions for the \texttt{FormTracer} Mathematica package. Information on these can be found by calling

{
\vspace{1ex}
\footnotesize
\begin{mmaCell}{Input}
\mmaDef{TBInfo}["Extensions"]
\end{mmaCell}
}

\noindent
A list of all \texttt{FormTracer}-known definitions can be accessed by

{
\vspace{1ex}
\footnotesize
\begin{mmaCell}{Input}
\mmaDef{TBInfo}["FormTracer"]
\end{mmaCell}
\noindent
}

\noindent
For a concrete example on how to use the package with the quark-gluon vertex as an application, see also \Cref{app:Mathem}.

\section{Summary \& Conclusions}
\label{sec:conclusions}

In this work, we have given a comprehensive introduction to the principles governing the construction of tensor bases and the associated projection operators, taking into account the full structure of a given interaction. 
With this setup at hand, we have discussed the topic of basis optimisation in \Cref{sec:optimisation} in depth. Specifically, we have related two points of view with each other: Optimisation on the level of the chosen basis' metric optimisation on the level of flow equations, see \Cref{sec:optimisation}. 

As we have pointed out, it is crucial for the optimisation of momentum-dependent tensor bases to avoid momentum-dependent irregularities in the basis choice. These irregularities are encoded in zeros or singularities of the associated metric $g_{ij}$. Using the quark-gluon vertex as an example, we have also shown in \Cref{sec:momenta} how such irregularities can be systematically removed. 
Moreover, the conceptual and computational advances have been illustrated at the example of the four-fermion vertex in \Cref{sec:fourFermiVertex}.
An explicit numerical investigation of flow optimisation as presented in \Cref{sec:optimisation} in the context of four-fermion interactions will be presented elsewhere.

The accompanying open-source Mathematica package \href{https://github.com/satfra/TensorBases.git}{\TensorBases} for the handling of projectors and interaction tensor bases is available on GitHub.
In addition to providing all the linear algebra routines explained in \Cref{sec:Preliminaries} and \Cref{sec:inner_product}, one can also restrict existing bases or construct new bases. 
It automatically reduces over-complete bases and computes all objects of interest in this respect, such as the metric, vertices and projection operators.
With this package, we also provide an extensible library of commonly used interaction bases.

\begin{acknowledgments}
	We thank Laura Classen, Chuang Huang, Manuel Reichert, Fabian Rennecke, Benedikt Schallmo, Moritz Thies, Jonas Turnwald, and Jonas Wessely for discussions. This work is done within the fQCD collaboration \cite{fQCD} and we thank its members for discussions and collaborations on related projects.
	FRS acknowledges funding by the GSI Helmholtzzentrum f\"ur Schwerionenforschung and by HGS-HIRe for FAIR.
	JB and NW acknowledge support by the Deutsche Forschungsgemeinschaft (DFG, German Research Foundation) – Project number 315477589 – TRR 211 and by the State of Hesse within the Research Cluster ELEMENTS (Project ID 500/10.006).
	Moreover, JB and AG acknowledge support by the Deutsche Forschungsgemeinschaft (DFG, German Research Foundation) through the Collaborative Research Center CRC 1245 “Nuclei: From Fundamental Interactions to Structure and Stars” – project number 279384907 – SFB 1245. This work is funded by the Deutsche Forschungsgemeinschaft (DFG, German Research Foundation) under Germany’s Excellence Strategy EXC 2181/1 - 390900948 (the Heidelberg STRUCTURES Excellence Cluster) and the Collaborative Research Centre SFB 1225 (ISOQUANT). It is also supported by EMMI.

\end{acknowledgments}

\appendix

\section{Notation}
\label{app:notation}

In the following, we briefly detail our notation for describing tensor bases of general QFTs.
To keep the notation concise and general, we collect all fields of the theory at hand into a single superfield~$\Phi$. 
We do this for all general arguments but make fields and indices explicit for specific examples.
Furthermore, we absorb all indices of the superfield, including momenta and group indices, into a single general multi-index.
	
An index~$a$ for a field~$\Phi^a$ contains momentum and possibly Lorentz, colour, flavour or further indices of the corresponding field.
Wherever indices are explicitly given, we use Greek letters~$\mu$,~$\nu$,~$\rho$,~$\sigma$,~$\dotsc$ for Lorentz indices and Latin letters~$a$,~$b$,~$c$,~$\dotsc$ for indices associated with any other type of group. 

While for bosons no additional structure needs to be imposed on the superfield, it is necessary to take into account that fermion and anti-fermion fields always come in pairs.
To that end, we use the field-space metric given by
\begin{equation}
	\gamma_{ab} = \gamma^{ab} = \begin{cases}
		\begin{pmatrix}
			0 & -1 \\
			1 & 0
		\end{pmatrix}\,\delta_{ab}
		& \text{if $a$ and $b$ are fermionic,} \\[3ex]
		\delta_{ab} & \text{if $a$ and $b$ are bosonic,}   \\[2ex]
		0               & \text{otherwise} 
		\,.
	\end{cases}
\end{equation}
One can now raise and lower indices of a super-field using the metric
\begin{equation}
	\Phi_a = \Phi^b\gamma_{ba} \,,
	\quad \quad
	\Phi^a = \gamma^{ab}\Phi_b \,.
\end{equation}
Raising and lowering indices, as introduced at the example of the superfield (vector), also applies to general higher rank tensors, e.g.
\begin{equation}
	M^{ab}_{\phantom{ab}c} = \gamma^{aa'} \gamma^{bb'} M_{a'b'}^{\phantom{a'b'}c'} \gamma_{c'c}
	\,.
\end{equation}
With this, derivatives of an arbitrary functional~$F[\Phi]$ are written as
\begin{equation}
	F_{a_1 a_2 \ldots a_n}[\Phi] =
	\frac{\delta}{\delta \Phi^{a_1}} \frac{\delta}{\delta \Phi^{a_2}} \ldots \frac{\delta}{\delta \Phi^{a_n}} F[\Phi]
	\,.
	\label{eq:Falpha}
\end{equation}
For convenience, we always consider all momenta to be incoming, which also fixes our Fourier convention:
\begin{equation}
	\Phi^a(x) = \int \frac{d^d p}{(2\pi)^d} e^{ipx}\Phi^a(p)
	\,.
\end{equation}
Finally, we write the general decomposition of an~$n$-th derivative of~$F[\Phi]$ as
\begin{equation}\label{eq:general_decomposition}
	F_{\alpha}[\Phi] = (2\pi)^d\ \delta^{(d)}\left( \sum_{i=1}^n p_i \right)\,
	\sum_{i=1}^{N_\alpha} \tau_{i,\alpha} \, \lambda_{i,\alpha}
	\,,
\end{equation}
where we have employed an even more compact notation by introducing the multi-index~$\alpha = (a_1, a_2, \ldots, a_n)$. 
This index includes all field indices.
Here, the~$\{\tau_{i,\alpha}\}$ are some basis of dimension~$N_\alpha$ of the tensor space of the Green's function, while the~$\{\lambda_{i,\alpha}\}$ are the coefficients of the expansion of~$F^\alpha$ within this basis.
	
Note that no summation over $\alpha$ is implied in \labelcref{eq:general_decomposition}: the basis elements may fully depend on the elements of the multi-index $\alpha$, whereas the coefficients depend at least on the momenta of the particles  contained in $\alpha$.
If additional ``continuous indices" are present, these can also be included as dependences of the dressings. 

For example, in \labelcref{eq:general_decomposition} the coefficients~$\lambda_i$ can be chosen such that they have an additional field dependence, e.g. coming from some composite operator $\phi[\Phi]$, which is useful if the associated symmetry is spontaneously broken. 
An example would be the composite field $\phi[\Phi] = \bar q q$ in QCD, which obtains a finite expectation value in the vacuum.
Any such rewriting either changes~$N_\alpha$ from uncountable infinity to a finite number, or, as in the above example, changes the expansion point of the interaction (i.e. the field background).
	
Furthermore, note that if~$F[\Phi] = \Gamma[\Phi]$, we call~$\Gamma_\alpha[\Phi]$ a \textit{vertex} of the theory and~$\{\lambda_i\}$ are the corresponding \textit{dressings}.

\section{Useful formulas for momentum parametrisations}
		
Using hyper-spherical coordinates, one can parametrise any four-momentum~$p$ using a radial variable $p_\tinytext{v}$, two cosines $z_1$, $z_2$, and an angle $\varphi$, as
\begin{equation}
	p = p_\tinytext{v} \begin{pmatrix}
		z_1 \\[1ex]
		\sqrt{1-z_1^2} z_2 \\[1ex]
		\sqrt{1-z_1^2} \sqrt{1-z_2^2} \cos{\varphi} \\[1ex]
		\sqrt{1-z_1^2} \sqrt{1-z_2^2} \sin{\varphi}
	\end{pmatrix}
	\,.
\end{equation}
For practical reasons, we also work out the relations between $S_0$, $a$, $s$, and the three incoming momenta $p_{1\ldots  3}$ in terms of scalar products:
\begin{equation}
	p_3^2 = (-p_1-p_2)^2 = \innerproduct{p_1}{p_1} + \innerproduct{p_2}{p_2} + 2 \innerproduct{p_1}{p_2}
	\,,
\end{equation}
and hence
\begin{align}
	\mathcal{S}_0 &= \frac{\sqrt{\innerproduct{p_1}{p_1} + \innerproduct{p_2}{p_2} + \innerproduct{p_1}{p_2}}}{\sqrt3} 
	\,,\notag\\[1ex]
	a &= \frac{\sqrt{3}}{2} \frac{\innerproduct{p_1}{p_1} + 2 \innerproduct{p_1}{p_2} }{\innerproduct{p_1}{p_1} + \innerproduct{p_2}{p_2} + \innerproduct{p_1}{p_2}}
	\,,\notag\\[1ex]
	s &= \frac{\innerproduct{p_1}{p_2} + \innerproduct{p_2}{p_2} - \frac{1}{2} \innerproduct{p_1}{p_1}}{\innerproduct{p_1}{p_1} + \innerproduct{p_2}{p_2} + \innerproduct{p_1}{p_2}}
	\,,
\end{align}
which can be inverted to arrive at
\begin{align}
	\innerproduct{p_1}{p_1} &= 2 (1-s) \mathcal{S}_0^2 
	\,,\notag\\[1ex]
	\innerproduct{p_2}{p_2} &= \left(2-\sqrt{3}a + s\right) \mathcal{S}_0^2
	\,,\notag\\[1ex]
	\innerproduct{p_1}{p_2} &= \left(-1+\sqrt{3}a+s\right) \mathcal{S}_0^2
	\,.
\end{align}
%

\section{Using the \TensorBases Mathematica package}
\label{app:Mathem}
In this appendix, we shall briefly introduce some of the features of the \TensorBases Mathematica package.
To get started, load the package after installation with 

{
\vspace{1ex}
\footnotesize
\begin{mmaCell}{Input}
Get["TensorBases`"]
\end{mmaCell}
}

\noindent
This will also print out some information on the package, including multiple strings on which you can use the \mmaInlineCell{Code}{TBInfo[...]} command to access the full documentation of the package.
On the other hand, one can always directly view the documentation strings using

{
\vspace{1ex}
\footnotesize
\begin{mmaCell}{Input}
TBInfo::\mmaStr{usage}
\end{mmaCell}
\begin{mmaCell}{Output}
TBInfo[_String]
Return information on a given object. 
TBInfo[] prints all available bases with some usage information.
TBInfo[BasisName] prints detailed information provided by this basis.
TBInfo["FormTracer"] prints all defined groups and identites which FormTracer currently knows.
TBInfo["Extensions"] prints all extensions to FormTracer, defined by the TensorBases package.
TBInfo["Momenta"] prints all momentum transformations that can be performed by the TensorBases package.
\end{mmaCell}
}

\noindent
To see a list of all bases predefined in the \TensorBases database, one can directly use the \mmaInlineCell{Code}{TBInfo[]} command without any argument.
To access the basis elements of a certain tensor basis, one can use the \mmaInlineCell{Code}{TBGetBasisElement} command:

{
\vspace{1ex}
\footnotesize
\begin{mmaCell}[addtoindex=1]{Input}
TBGetBasisElement::\mmaStr{usage}
TBGetBasisElement["AqbqDirect",1,\{p1,mu,a\},\{p2,d2,A2,F2\},\{p3,d3,A3,F3\}]
\end{mmaCell}
\begin{mmaCell}{Output}
TBGetBasisElement[BasisName_String,n_Integer,indices___]
Obtains the n-th element of the specified basis. 
The given indices must match the ones specified by the basis, see TBInfo[].
If no indices are given, the standard indices specified by the basis are used.
TBGetBasisElement[BasisName_String,All,indices___]
Returns a list with all elements of the specified basis. 
The given indices must match the ones specified by the basis, see TBInfo[].
If no indices are given, the standard indices specified by the basis are used.
\end{mmaCell}
\begin{mmaCell}{Output}
i deltaFundFlav[F2,F3] gamma[rho$10648,d2,d3] TCol[a,A2,A3] transProj[-p2-p3,mu,rho$10648]
\end{mmaCell}
}

\noindent
It is also possible to drop the indices when this function is called. 
In this case, a set of standard indices will be used in the output.
Every basis provides their inner product through a function that returns an operator which can be used with \mmaInlineCell{Code}{TBGetBasisElement}:

{
\vspace{1ex}
\footnotesize
\begin{mmaCell}{Input}
TBGetInnerProduct::\mmaStr{usage}
TBGetInnerProduct["AqbqDirect"][TBGetBasisElement, 1, TBGetBasisElement, 1]//FormTrace//Simplify
\end{mmaCell}
\begin{mmaCell}{Output}
TBGetInnerProduct[BasisName_String]
Returns the bilinear operator \(\mathcal{O}\) that represents the inner product of the specified basis. 
It can be called as \(\mathcal{O}\)[Tensor1, n, Tensor2, m], where Tensor1 and Tensor2 are functions with signatures
Tensor[BasisName_String, n_Integer, indices___].
For example, \(\mathcal{O}\)[TBGetBasisElement, 1, TBGetBasisElement, 1] returns <\mmaSub{e}{i},\mmaSub{e}{j}>.
\end{mmaCell}
\begin{mmaCell}{Output}
-6 (-1+\mmaSup{Nc}{2}) Nf
\end{mmaCell}
}

\noindent
To directly access the metric, one can use the \mmaInlineCell{Code}{TBGetMetric} command and the inverse metric with \mmaInlineCell{Code}{TBGetInverseMetric}:

{
\vspace{1ex}
\footnotesize
\begin{mmaCell}{Input}
TBGetMetric::\mmaStr{usage}
TBGetMetric["AqbqDirect"][[1,1]]
\end{mmaCell}
\begin{mmaCell}{Output}
TBGetMetric[BasisName_String]
Returns the metric of the specified basis, i.e. the matrix \mmaSub{g}{ij} = <\mmaSub{e}{i},\mmaSub{e}{j}>, where the \mmaSub{e}{i} are the basis
elements of the basis.
\end{mmaCell}
\begin{mmaCell}{Output}
-6 (-1+\mmaSup{Nc}{2}) Nf
\end{mmaCell}
	
\begin{mmaCell}{Input}
TBGetInverseMetric::\mmaStr{usage}
TBGetInverseMetric["AqbqDirect"][[1,1]]
\end{mmaCell}
\begin{mmaCell}{Output}
TBGetInverseMetric[BasisName_String]
Returns the inverse of the metric of the specified basis, i.e. the matrix \mmaSup{\mmaSub{g}{ij}}{-1} = (<\mmaSub{e}{i},\mmaSub{e}{j}>\mmaSup{)}{-1}, where the \mmaSub{e}{i}
are the basis elements of the basis.
\end{mmaCell}
\begin{mmaCell}{Output}
\mmaFrac{1}{6 Nf-6 \mmaSup{Nc}{2} Nf}
\end{mmaCell}
}

\noindent
Most importantly, one obtains the projection operators of any basis using the \mmaInlineCell{Code}{TBGetProjector} command, which works just like the \mmaInlineCell{Code}{TBGetBasisElement} command introduced before,

{
\vspace{1ex}
\footnotesize
\begin{mmaCell}{Input}
TBGetProjector::\mmaStr{usage}
TBGetProjector["AqbqDirect",1,\{p1,mu,a\},\{p2,d2,A2,F2\},\{p3,d3,A3,F3\}]
TBGetInnerProduct["AqbqDirect"][TBGetProjector, 1, TBGetBasisElement, 1]//FormTrace//Simplify
TBGetInnerProduct["AqbqDirect"][TBGetProjector, 1, TBGetBasisElement,8]//FormTrace//Simplify
\end{mmaCell}
\begin{mmaCell}{Output}
TBGetBasisProjector[BasisName_String,n_Integer,indices___]
Returns the n-th projector, which is defined by \mmaSup{\mmaSub{g}{nj}}{-1}\mmaSub{e}{j}.
The given indices must match the ones specified by the basis, see TBInfo[]. 
If no indices are given, the standard indices specified by the basis are used.
TBGetVertex[BasisName_String,All,indices___]
Returns a list with all projectors of the specified basis, defined by \mmaSup{\mmaSub{g}{nj}}{-1}\mmaSub{e}{j}.
The given indices must match the ones specified by the basis, see TBInfo[].
If no indices are given, the standard indices specified by the basis are used.
\end{mmaCell}
\begin{mmaCell}{Output}
\mmaFrac{i deltaFundFlav[F2,F3] gamma[rho$10775,d2,d3] TCol[a,A2,A3] transProj[-p2-p3,mu,rho$10775]}{6 Nf-6 \mmaSup{Nc}{2} Nf}
\end{mmaCell}
\begin{mmaCell}{Output}
1
\end{mmaCell}
\begin{mmaCell}{Output}
0
\end{mmaCell}
}

\noindent
Suppose you wish to restrict this basis to the most important elements 1, 4 and 7. This can be done by using 

{
\vspace{1ex}
\footnotesize
\begin{mmaCell}{Input}
TBRestrictBasis::usage
TBRestrictBasis["AqbqDirect","AqbqDirect147",\{1,4,7\}]
\end{mmaCell}
\begin{mmaCell}{Output}
TBRestrictBasis[inBasisName_String, outBasisName_String, {indices__Integer}, CacheDirectory_String:"./TBCache"]
Restrict an existing basis. The new basis will be called outBasisName and only contain the basis
elements specified by the given indices.
\end{mmaCell}
}

\noindent
This will create a new basis called ``AqbqDirect147" and stores it in the folder \texttt{./TBCache}, where the filename is constructed from the arguments of the command \mmaInlineCell{Code}{TBRestrictBasis}.
This new basis has all functions available that the original basis also had, but calculated from the new, restricted metric.
Bases generated with the package can be stored in any directory. For example, the basis from the previous example can be stored by

{
\vspace{1ex}
\footnotesize
\begin{mmaCell}{Input}
TBExportBasis::usage
TBExportBasis["AqbqDirect147", "./MyBasisCollection/"]
\end{mmaCell}
\begin{mmaCell}{Output}
TBExportBasis[BasisName_String,folder_String:"./"]
Export a basis definition file. The basis with the name BasisName has to be loaded in memory. 
If the optional argument folder is given, this will be the location where the exported basis definition
will be placed.
\end{mmaCell}
}

\noindent
To load it again, one can use \mmaInlineCell{Code}{TBImportBasis}. To avoid an error, we will first un-register the basis and then reload it:

{
\vspace{1ex}
\footnotesize
\begin{mmaCell}{Input}
TBUnregister::usage
TBUnregister["AqbqDirect147"]
\end{mmaCell}
\begin{mmaCell}{Output}
TBUnregister[BasisName_String]
Remove an existing basis from internal memory. This does not delete or change any files on disk.
\end{mmaCell}
\begin{mmaCell}{Input}
TBImportBasis::usage
TBImportBasis["./MyBasisCollection/AqbqDirect147.m"]
\end{mmaCell}
\begin{mmaCell}{Output}
TBImportBasis[BasisDefinitionFile_String,CacheDirectory_String:"./TBCache"]
Import a custom basis definition file. 
The optional argument CacheDirectory can be set to choose a specific location where the intermediate files
from processing the basis are stored.
\end{mmaCell}
}

\noindent
Now, suppose we wish to define a completely new basis directly from a notebook. This can be done by calling

{
\vspace{1ex}
\footnotesize
\begin{mmaCell}{Input}
TBConstructBasis[
	"FourFermionBasis",
	\{\},
	\mmaOver{\(\pmb{\psi}\)}{_}\(\pmb{\psi}\)\mmaOver{\(\pmb{\psi}\)}{_}\(\pmb{\psi}\),
	2(Tensor[1,2,3,4]-Tensor[1,4,3,2]),
	2Tensor1[1,2,3,4](Tensor2[2,1,4,3]-Tensor2[4,1,2,3]),
	"Built within Mathematica",
	"Franz R. Sattler",
	"...",
	\{\{p1,d1\},\{p2,d2\},\{p3,d3\},\{p4,d4\}\},
	\{p4->-p3-p2-p1\},
	\{\{
		deltaDirac[d1,d2]deltaDirac[d3,d4],
		gamma[mu,d1,d2]gamma[mu,d3,d4],
		gamma5[d1,d2]gamma5[d3,d4],
		gamma[mu,d1,dint1]gamma5[dint1,d2]gamma[mu,d3,d3int]gamma5[d3int,d4],
		sigma[mu,nu,d1,d2]sigma[mu,nu,d3,d4]
	\}\},
	\{\}
];
\end{mmaCell}
}

\noindent
Here, we have constructed a four-fermion basis by specifying the inner product, the vertex structure, all indices, the momentum conservation rule and a list of tensors. \TensorBases constructs a metric from the list of given tensors and reduces it using the given scalar product until the resulting metric has full rank.
More information on this command can be obtained by calling \mmaInlineCell{Code}{TBInfo["BaseBuilder"]} or \mmaInlineCell{Code}{TBConstructBasis::usage}.

After having constructed the basis, it is available in the current notebooks's \TensorBases database. With the above example, the resulting basis has three elements, and the two last tensors have been discarded as they are combinations of the first three elements:

{
\vspace{1ex}
\footnotesize
\begin{mmaCell}{Input}
TBGetMetric["FourFermionBasis"]//MatrixForm
TBGetBasisElement["FourFermionBasis",1]
TBGetBasisElement["FourFermionBasis",2]
TBGetBasisElement["FourFermionBasis",3]
\end{mmaCell}
\begin{mmaCell}[form=MatrixForm]{Output}
(24	-32	-8
 -32	192	32
 -8	32	24)
\end{mmaCell}
\begin{mmaCell}{Output}
deltaDirac[d1,d2] deltaDirac[d3,d4]
\end{mmaCell}
\begin{mmaCell}{Output}
gamma[mu$17896,d1,d2] gamma[mu$17896,d3,d4]
\end{mmaCell}
\begin{mmaCell}{Output}
gamma5[d1,d2] gamma5[d3,d4]
\end{mmaCell}
}

\noindent
We may again export the basis, i.e., the metric, inverse metric, projectors and vertices. Furthermore, one can also write the cache to disk, which will reduce the runtime for any future import thereof:

{
\vspace{1ex}
\footnotesize
\begin{mmaCell}{Input}
TBExportCache::\mmaStr{usage}
TBExportBasis["FourFermionBasis"]
TBExportCache["FourFermionBasis"]
\end{mmaCell}
\begin{mmaCell}{Output}
TBExportCache[BasisName_String,CacheFolder_String:"./TBCache"]
Exports everything in memory of the Basis BasisName onto disk in the folder CacheFolder.
\end{mmaCell}
}

\noindent
We conclude by noting that \TensorBases provides a cache for all bases which are pre-defined in its database, which allows to provide them quickly and takes much less time than fully re-building the basis cache each time one imports the \TensorBases package.

%

\section{Fierz Identities}
\label{app:fierzIdentities}

Fierz identities allow us to rewrite bilinears of the product of two fermionic spinors as a linear combination of products of the bilinear of spinors. In general, these can be written as a set of matrix identities that arise from interchanging certain indices.
In this appendix, we shall derive three Fierz transformations that have been used in the present work.

For this purpose, let us define~$\lbrace \mathcal F^{(i)} \rbrace$ to be a basis of a matrix space, which fulfils the following properties
\begin{equation}\label{eq:orthogonality_completeness}
	\textrm{Tr} \left( \mathcal F^{(i)} \mathcal F^{(j)} \right)
	\equiv \sum_{ a_1  a_2} \mathcal F^{(i)}_{ a_1,  a_2} \mathcal F^{(j)}_{ a_2  a_1}
	= \delta^{ij} \qquad \text{and} \qquad \sum_j \mathcal F^{(j)}_{ a_1 a_2} \mathcal F^{(j)}_{ a_3 a_4} = \delta_{ a_1 a_4}\delta_{ a_2 a_3}
	\,.
\end{equation}
which we refer to be the orthogonality and completeness relation, respectively.
Upper indices denote the basis elements, while lower indices denotes the matrix entries that can in principle be a collective index set for, e.g. colour, flavour and Dirac space.
Supposing that~$\lbrace \mathcal F^{(i)} \rbrace$ spans the whole matrix space under consideration, each matrix~$M$ in this space can be expressed as a linear combination of the basis elements~$\lbrace \mathcal F^{(i)} \rbrace$, i.e.
\begin{equation}\label{eq:matrix_expansion}
	M = \sum_i \textrm{Tr} \left(M \mathcal F^{(i)} \right) \mathcal F^{(i)} 
	\,.
\end{equation}
With this at hand, we can now derive the Fierz identities.

\subsection{Fierz Identity I}
\label{app:fierzIdentityI}

Let us consider two matrices~$M$ and~$\tilde M$. We write
\begin{equation}\label{eq:fierzI}
	M_{ a_1 a_2} \tilde M_{ a_3 a_4}
	= \sum_{nmpq} (\delta_{ a_1 b_1} \delta_{ a_4 b_4}) (\delta_{ a_2 b_2} \delta_{ a_3 b_3})  M_{b_1b_2} \tilde M_{b_3 b_4}
	\overset{\eqref{eq:orthogonality_completeness}}{=} \sum_{ij} \textrm{Tr} \left( \mathcal F^{(i)} M \mathcal F^{(j)} \tilde M \right) \mathcal F^{(i)}_{ a_1 a_4} \mathcal F^{(j)}_{ a_3 a_2} \, ,
\end{equation}
where we have again used the completeness relation for the Kronecker deltas in parentheses in the last step.
Note that, loosely speaking, this Fierz transformation interchanges indices~$a_2$ and~$ a_4$.
Attaching fermionic fields to \labelcref{eq:fierzI} yields
\begin{equation}\label{eq:fierzI_fermions}
	(\bar \psi M \psi) (\bar \psi \tilde M \psi) = - \sum_{ij} \textrm{Tr} \left( \mathcal F^{(i)} M \mathcal F^{(j)} \tilde M \right) (\bar \psi \mathcal F^{(i)} \psi) (\bar \psi \mathcal F^{(j)} \psi)
	\,,
\end{equation}
where the minus sign origins from the Grassman nature of the fermionic fields.

\subsection{Fierz Identity II}
\label{app:fierzIdentityII}

Let us again consider two matrices~$M$ and~$\tilde M$.
We write
\begin{equation}\label{eq:fierzII}
	M_{ a_1 a_2} \tilde M_{ a_3 a_4}
	= \sum_{b_1b_2b_3b_4} (\delta_{ a_1b_1}  \delta_{ a_3b_3}) (\delta_{ a_2b_2}\delta_{ a_4b_4})  M_{b_1b_2} \tilde M_{b_3b_4}
	\overset{\eqref{eq:orthogonality_completeness}}{=} \sum_{ij} \textrm{Tr} \left( \mathcal F^{(i)} M \mathcal F^{(j)} \tilde M^T \right) \mathcal F^{(i)}_{ a_1 a_3} \mathcal F^{(j)}_{ a_4 a_2} 
	\,,
\end{equation}
where we have used the completeness relation for the Kronecker deltas in parentheses in the last step.	
By attaching fermionic fields to \labelcref{eq:fierzII}, we obtain
\begin{equation}\label{eq:fierzII_fermions}
	(\bar \psi M \psi) (\bar \psi \tilde M \psi) = \sum_{ij} \textrm{Tr} \left( \mathcal F^{(i)} M \mathcal F^{(j)} \tilde M^T \right) (\bar \psi \mathcal F^{(i)} \bar \psi^T) (\psi^T \mathcal F^{(j)} \psi)
	\,,
\end{equation}
Loosely speaking, this Fierz transformation transforms a four-fermion basis element from the standard from $(\bar \psi \dots \psi)(\bar \psi \dots \psi)$ into a linear combination of the form $(\bar \psi \dots \bar \psi^T)(\psi^T \dots \psi)$.
Note that the colour-superconducting channel is typically written in the latter form, namely $\mathcal{L}_\textrm{csc} = (\bar\psi \gamma_5 \mathcal{C} \tau_2 i \epsilon_a \bar\psi^T)(\psi^T \mathcal{C}\gamma_5 \tau_2 i \epsilon_a \psi)$.

\subsection{Fierz Identity III}
Finally, let us introduce a third Fierz identity. It is the ``inverse" of the one introduced in \Cref{app:fierzIdentityII}. We start with
\begin{equation}\label{eq:fierzIII}
	M_{ a_1 a_2} \tilde M_{ a_3 a_4}
	= \sum_{b_1b_2b_3b_4} (\delta_{ a_1b_1}  \delta_{ a_4b_4}) (\delta_{ a_2b_2}\delta_{ a_3b_3})  M_{b_1b_2} \tilde M_{b_3b_4}
	\overset{\eqref{eq:orthogonality_completeness}}{=} \sum_{ij} \textrm{Tr} \left( \mathcal F^{(i)} M \left(\mathcal F^{(j)}\right)^T \tilde M \right) \mathcal F^{(i)}_{ a_1 a_4} \mathcal F^{(j)}_{ a_2 a_3} 
	\,.
\end{equation}
Attaching fermionic fields to \labelcref{eq:fierzIII}, we eventually find 
\begin{equation}\label{eq:fierzIII_fermions}
	(\bar \psi M \bar\psi^T) (\psi^T \tilde M \psi) = \sum_{ij} \textrm{Tr} \left( \mathcal F^{(i)} M \left(\mathcal F^{(j)}\right)^T \tilde M \right) (\bar \psi \mathcal F^{(i)} \psi) (\bar \psi \mathcal F^{(j)}\psi)
	\,.
\end{equation}

\subsection{Quark Current Bases}
\noindent
The Dirac space is spanned by 16 elements. We choose
\begin{equation}\label{eq:DiracBasis}
	\lbrace \mathcal F^{(i)} \rbrace = \lbrace
	\mathds{1}_\textrm{D},\, \gamma_\mu, \sigma_{\mu\nu},\, \imag \gamma_\mu \gamma_5,\, \gamma_5\rbrace 
	\,.
\end{equation}
Here,~$\gamma_\mu$ are the Dirac matrices, $\sigma_{\mu\nu} = \frac{\imag}{2}[\gamma_\mu,\gamma_\nu]$, and~$\gamma_5$ is the chiral Dirac matrix.
Alternatively, one can use a basis inspired by the structure of the colour-superconducting channel~$\mathcal{L}_\textrm{csc}$, namely
\begin{align}
	\lbrace \mathcal F^{(i)}_\Delta \rbrace = \lbrace \mathcal{C},\, \imag \gamma_0 \mathcal{C},\, \gamma_1 \mathcal{C},\, \imag \gamma_2 \mathcal{C},\, &\gamma_3 \mathcal{C},\,
    \imag \sigma_{01} \mathcal{C},\, \sigma_{02} \mathcal{C},\, \imag \sigma_{03} \mathcal{C},
    \imag \sigma_{12} \mathcal{C},\, \sigma_{13} \mathcal{C},\, 
    \notag\\[1ex]
    &\imag \sigma_{23} \mathcal{C},\,
    \imag \gamma_0 \gamma_5 \mathcal{C},\, \gamma_1 \gamma_5 \mathcal{C},\,
    \imag \gamma_2 \gamma_5 \mathcal{C},\, \gamma_3 \gamma_5 \mathcal{C},\, \gamma_5 \mathcal{C}
	\rbrace
	\,.
\end{align}
%

\section{Fierz Identity for the Colour-Superconducting Channel}
\label{app:fierzIdentitieCSC}
\noindent
Using~\labelcref{eq:fierzIII_fermions}, the Fierz-transformation of the colour-superconducting channel is given by
\begin{align}
	&(\bar\psi \gamma_5 \mathcal{C} \tau_2 i \epsilon_a \bar\psi^T)(\psi^T \mathcal{C}\gamma_5 \tau_2 i \epsilon_a \psi)
	\nonumber\\
	& =
	\frac{1}{12}\left( (\bar\psi\psi)^2 + (\bar\psi\gamma_\mu\psi)^2 + (\bar\psi\gamma_5\psi)^2  - \frac{1}{2}(\bar\psi\sigma_{\mu\nu}\psi)^2 - (\bar\psi i \gamma_\mu\gamma_5\psi)^2 \right)
	\nonumber\\
	& -
	\frac{1}{12}\left( (\bar\psi\tau_i\psi)^2 + (\bar\psi\gamma_\mu\tau_i\psi)^2 + (\bar\psi\gamma_5 \tau_i \psi)^2  - \frac{1}{2}(\bar\psi\sigma_{\mu\nu} \tau_i \psi)^2 - (\bar\psi i \gamma_\mu\gamma_5 \tau_i \psi)^2  \right)
	\nonumber\\
	& -
	\frac{1}{4}\left( (\bar\psi T^a\psi)^2 + (\bar\psi\gamma_\mu T^a \psi)^2 + (\bar\psi\gamma_5 T^a\psi)^2  - \frac{1}{2}(\bar\psi\sigma_{\mu\nu}T^a \psi)^2 - (\bar\psi i \gamma_\mu\gamma_5 T^a \psi)^2 \right)
	\nonumber\\
	& +
	\frac{1}{4}\left( (\bar\psi T^a\tau_i\psi)^2 + (\bar\psi\gamma_\mu T^a \tau_i \psi)^2 + (\bar\psi\gamma_5 T^a \tau_i \psi)^2  - \frac{1}{2}(\bar\psi\sigma_{\mu\nu}T^a \tau_i \psi)^2 - (\bar\psi i \gamma_\mu\gamma_5 T^a \tau_i \psi)^2\right)
	\,.
	\label{eq:cscFierzTrafo}
\end{align}
Here, $\gamma_\mu$ are the Dirac matrices, $\sigma_{\mu\nu} = \frac{\imag}{2}[\gamma_\mu,\gamma_\nu]$ and $\epsilon_a$ are the antisymmetric generators of the~SU$(3)$ colour transformations.
Furthermore, $T^a_c$ and~$\tau_i$ are the fundamental representations of the generators of the colour gauge group and the SU$(N_f)$, respectively, and $\mathcal{C}$ is the charge conjugation matrix.
Sums over repeated indices are implied and unity matrices are suppressed.

\newpage
\bibliography{bibliography}

\end{document}